# The Generative Reasonable Person

*Yonathan A. Arbel*[*]

This Article introduces the *generative reasonable person*, a new tool for estimating how ordinary people judge reasonableness. As claims about AI capabilities often outpace evidence, the Article proceeds empirically: adapting randomized controlled trials to large language models, it replicates three published studies of lay judgment across negligence, consent, and contract interpretation, drawing on nearly 10,000 simulated decisions.

The findings reveal that models can replicate subtle patterns that run counter to textbook treatment. Like human subjects, models prioritize social conformity over cost-benefit analysis when assessing negligence, inverting the hierarchy that textbooks teach. They reproduce the paradox that material lies erode consent less than lies about a transaction's essence. And they track lay contract formalism, judging hidden fees more enforceable than fair.

For two centuries, scholars have debated whether the reasonable person is empirical or normative, majoritarian or aspirational. But much of this debate assumed a constraint that no longer holds: that lay judgments are expensive to surface, slow to collect, and unavailable at scale. Generative reasonable people loosen that constraint. They offer judges empirical checks on elite intuition, give resource-constrained litigants access to simulated jury feedback, and let regulators pilot-test public comprehension, all at a fraction of survey costs. The reasonable person standard has long functioned as a vessel for judicial intuition precisely because the empirical baseline was missing. With that baseline now available, departures from lay understanding become transparent rather than hidden, a choice to be justified, not a fact to be assumed. Properly cabined, the generative reasonable person may become a dictionary for reasonableness judgments.

---

* Professor of Law, University of Alabama, School of Law, Director AI Legal Studies. I am grateful for the valuable feedback and insights provided by J. Shahar Dillbary, Niva Elkin-Koren, Andres Swaicki, David A. Hoffman, Christopher Brett Jaeger, Ben McMichael, Peter N. Salib, Roseanna Sommers, Kevin Tobia, Andrew Coan, and Matthew Tokson. This work also benefited from comments by participants in the NYU Empirical Contracts Workshop, Tel Aviv University Law & Tech Seminar, and faculty workshops in Miami and Arizona Law. Justin Heydt and Andrew Robitaille provided important research assistance. Any errors are my own.

## INTRODUCTION

In the contracts chestnut *Leonard v. Pepsico*, the case hinged on a single question: how would people reasonably interpret a television ad offering prizes for Pepsi points, among them a fighter jet?[1] Judge Wood emerged from her chambers with an answer: "no reasonable, objective person," she wrote, "would have understood the commercial to be an offer."[2] With that, she dismissed the claim.

The case always stirs some controversy in the classroom. But ask yourself not whether you agree with her decision, but rather what tools did the law give judges, alone in chambers, to read into the minds of the teenagers Pepsi's ad targeted? Intuitive judgments are hard. Judges are rarely the target audience for the messages they are asked to interpret, and developmental psychology and neuroscience suggest that adolescents are especially sensitive to social-evaluative and reward cues found in advertising.[3] Yet for centuries, determinations of reasonable interpretations have demanded exactly this mind-reading. Critics charge that such unguided standards risk realizing legal realists' greatest fear: a vessel for judicial bias masquerading as common sense.[4]

For two centuries, we have lacked a cheap, scalable, empirical method for making lay reasonableness judgments legible to legal institutions.[5] Jury trials were designed to channel community standards, but they suffer from selection effects,

---

[1] Leonard v. PepsiCo, Inc., 88 F. Supp. 2d 116, 127 (S.D.N.Y. 1999), aff'd, 210 F.3d 88 (2d Cir. 2000). Keith A. Rowley, *You Asked for It, You Got It ... Toy Yoda: Practical Jokes, Prizes, and Contract Law*, 3 NEV. L.J. 526, 536 (2003) (finding that the case features in eight out of 15 sampled textbooks).

[2] *Id.* at 131.

[3] *See, e.g.,* Laurence Steinberg, *A Social Neuroscience Perspective on Adolescent Risk-Taking*, 28 DEV. REV. 78, 79–86, 96–100 (2008); Sarah-Jayne Blakemore, *The Social Brain in Adolescence*, 9 NAT. REV. NEUROSCI. 267, 267–72 (2008) (reviewing developmental changes in social-cognitive processing, including how social cues and social information are interpreted during adolescence); Peter H. Wright, Marian Friestad & David M. Boush, *The Development of Marketplace Persuasion Knowledge in Children, Adolescents, and Young Adults*, 24 J. PUB. POL'Y & MKTG. 222, 222–27 (2005). The FTC, for instance, consistently emphasizes that advertising to specific age groups should be evaluated based on different standards. See FTC Policy Statement on Deception, (Oct, 13 1983) ("When representations or sales practices are targeted to a specific audience, such as children, the elderly, or the terminally ill, the Commission determines the effect of the practice on a reasonable member of that group."); FTC, Guides Concerning the Use of Endorsements and Testimonials in Advertising, 87 Fed. Reg. 44,288 (July 26, 2022) (request for comment) ("The Commission recognizes that it is difficult for children – especially younger children – to discern ads from entertainment.")

[4] The critical literature is vast. *See e.g.,* Leonardo J. B. Amorim, *Reasonable Interpretation: A Radical Legal Realist Critique*, 33 INT'L J. SEMIOT. L. 1043, 1056 (2020), ("radical realism allows the observer to notice that the appeal to "reasonability" functions as a joker in legal argumentation, a token allowing the interpreter and the public to cope with unconscious prejudgements, biases and external pressures."); David Zaring, *Rule by Reasonableness*, 63 ADMIN. L. REV. 525, 552 (2011) (noting that "One strong critique of reasonableness review in administrative law is that it would allow judges to enact their political preferences); MAYO MORAN, RETHINKING THE REASONABLE PERSON: AN EGALITARIAN RECONSTRUCTION OF THE OBJECTIVE STANDARD, 16-17 (2003) ("unsurprisingly, the reasonable person often turns out to bear a rather suspicious similarity to the judge."); Jeffrey M. Hayes, T*o Recuse or to Refuse: Self-Judging and the Reasonable Person Problem,* 33 J. LEGAL PROF. 85, 88 (2008) (arguing that judges make determinations "with a concept of "reason" that is uniquely shaped by their own environment, which means that in practice the so-called objective reasonable person standard collapses into subjectivity."); Audrey L. Cerfoglio, Emily M. Petrie, Monica K. Miller, *Is "Reasonable" Reasonable? A Content Analysis on Judges' Perceptions of the "Reasonable Person' Standard*, 57 UIC L. REV. 743, 747 (2024) (emphasizing inter-judge variability due to differences life experiences); *See also* Ain Simpson, Mark D. Alicke, Ellen Gordon & David Rose, *The Reasonably Prudent Person, or Me?,* 50 J. APPLIED SOC. PSYCHOL. 313 (2020) (Finding that people rely more heavily on their own projected behavior than on "reasonably prudent person" standards when judging others' harmful actions); Lawrence Solan, Terri Rosenblatt & Daniel Osherson, Essay, *False Consensus Bias in Contract Interpretation*, 108 COLUM. L. REV. 1268, 1269 (2008)

[5] Both statutory and constitutional analysis are beyond the scope of the analysis here, but it is worth noting the growing jurisprudential import of "ordinary" meaning. *See* Bostock v. Clayton Cnty., 140 S. Ct. 1731, 1738 (2020). *See also* Jesse M. Cross, The Fair Notice Fiction, 75 Ala. L. Rev. 487, 488-9 (2023) ("the Court increasingly would prioritize a single concern: the original public meaning of statutes")

adversarial distortion, and inaccessibility for most disputes. [6] Surveys can measure public perception, but they are costly, manipulable, slow, and remain underutilized. [7] Mock juries and jury consultants are available only to wealthy litigants. [8] Some have responded by declaring the reasonable person a purely normative construct,[9] untethered from empirical reality. But it is worth asking how much of this is principled jurisprudence and how much is coping with our lack of tools. [10] At the end, in most cases all we have is judicial intuition, educated

---

[6] Shamena Anwar, Patrick Bayer, & Randi Hjalmarsson, *The Impact of Jury Race in Criminal Trials*, 127 Q. J. ECON. 1017 (2012). The need for surveys was highlighted by Chief Justice Roberts' comments during oral argument in a statutory interpretation case: "[O]ur objective is to settle upon the most natural meaning of the statutory language to an ordinary speaker of English, right? . . . So the most probably useful way of settling all these questions would be to take a poll of 100 ordinary speakers of English and ask them what [the statute] means, right?" Transcript of Oral Argument at 51-52, Facebook, Inc. v. Duguid, 141 S. Ct. 1163 (2021) (No. 19-511). *Cited in* Kevin Tobia et. al., *Ordinary Meaning and Ordinary People*, 171 U. PA. L. REV. 365, 371 (2023).

[7] *See* Omri Ben-Shahar & Lior J. Strahilevitz, *Interpreting Contracts via Surveys and Experiments*, 92 N.Y.U. L. Rev. 1753 (2017). *See also* Thomas R. Lee & Stephen C. Mouritsen, *Judging Ordinary Meaning*, 127 YALE L.J. 788 (2018). By one estimate, each survey costs the federal government $65,000, Federal Trade Commission, *Public Comment on Methodology and Research Design for Conducting a Study of the Effects of Credit Scores and Credit-Based Insurance Scores on Availability and Affordability of Financial Products*, 69 FED. REG. 34,167 (June 18, 2004). Under the Paperwork Reduction Act, federal agencies must obtain approval from the Office of Management and Budget (OMB) before collecting information from ten or more non-federal persons, a process that requires internal agency review, certification, and a 60-day public comment period, often resulting in significant delays (44 U.S.C. §§ 3502(3), 3506-3507). *See also* Gisselle Bourns, Jennifer Nou & Stuart Shapiro, *Improving the Efficiency of the Paperwork Reduction Act*, REG. REV. (Oct. 30, 2018).

[8] Industry estimates of costs range from a few thousand dollars for a minimalistic version to upward of $50,000. *See* Casey Johnson, *Focus Groups on a Shoestring Budget*, Aitken, Aitken, & Cohn (Jul. 2, 2020), https://www.aitkenlaw.com/focus-groups-on-a-shoestring-budget ($8,000-$30,000); Merrie Pitera, What Does a Mock Trial Cost, IMS (Sep. 30, 2021), https://imslegal.com/articles/what-does-a-mock-trial-cost (or $10,000-$60,000); Andrew Guilford and Isabelle Ord, *Mocking Juries,* 18 CAL. LIT. 1 (2005) https://www.sheppardmullin.com/media/article/189_pub385.pdf ("a few thousand dollars on statistical information and jury consultation, or around $50,000 could be spent for a complete mock trial.").

[9] Scholars debate whether the reasonable person standard is descriptive, prescriptive, or a hybrid of both. The descriptive view treats reasonableness as reflecting actual societal norms. See In re Eastern Transp. Co. (The T.J. Hooper), 60 F.2d 737, 740 (2d Cir. 1932) (Hand, J.) ("In most cases reasonable prudence is in fact common prudence; but strictly it is never its measure."); Brian Sheppard, *The Reasonableness Machine*, 62 B.C. L. REV. 2259, 2288 (2021) (discussing the "Average Conduct Conception" of reasonableness). The prescriptive view argues that reasonableness is a normative ideal rather than an empirical observation. See Gregory C. Keating, *Reasonableness and Rationality in Negligence Theory*, 48 STAN. L. REV. 311, 339 (1996) (contending the standard embodies a community norm, not actual behavior); see also W. PAGE KEETON ET AL., PROSSER AND KEETON ON THE LAW OF TORTS 175 (5th ed. 1984) (describing the reasonable person as "a personification of a community ideal"). The hybrid approach recognizes both descriptive and prescriptive elements. See Alan Calnan, *The Nature of Reasonableness*, 105 CORNELL L. REV. ONLINE 81, 83 (2020) ("Some scholars say reasonableness is prescriptive, others say it reflects community values, and still others see it as a mix of both."); Kevin P. Tobia, *How People Judge What Is Reasonable*, 70 ALA. L. REV. 293, 296 (2018) (arguing reasonableness is "partly statistical and partly prescriptive"); Anita Bernstein, *The Communities That Make Standards of Care Possible*, 77 CHI.-KENT L. REV. 735, 740 (2002) (noting the standard's shifting balance between objective and subjective approaches). Even in his otherwise strong critique of experimental jurisprudence, Jimenez agrees that "some legal concepts—such as reasonableness—invite or require the use of lay understandings to determine at least part of their extension." Felipe Jimenez, *Some Doubts about Folk Jurisprudence: The Case of Proximate Cause,* U. CHI. L. REV. ONLINE 1 (2021))

[10] *See* Tobia, *supra* note 9, at 344 (highlighting the importance of understanding how ordinary people generate reasonableness judgments); Marvin L. Astrada & Scott B. Astrada, *Law, Continuity and Change: Revisiting the Reasonable Person Within the Demographic, Sociocultural and Political Realities of the Twenty-First Century*, 14 RUTGERS J.L. & PUB. POL'Y 200 (2017) (arguing that the rise in minority demographics demands reassessment of the reasonable person standard); Francesco Parisi & Georg von Wangenheim, *Legislation and Countervailing Effects from Social Norms*, *in* EVOLUTION AND DESIGN OF INSTITUTIONS 25, 29–30 (Christian Schubert & Georg v. Wangenheim eds., 2006) ( summarizing empirical studies that show that "[l]aws may more effectively influence behavioral outcomes when legal norms are aligned with the existing social values", whereas "[l]egitimacy is undermined when the content of the law departs from social norms").

guessing by precisely the elites criticized for being most removed from ordinary life.[11]

This Article provides such a tool. It develops a methodology for eliciting reasonableness judgments from large language models (LLMs), called "Silicon Randomized Controlled Trials" (s-RCT).[12] The approach treats independent model sessions as experimental subjects, leveraging their statelessness to enable controlled comparisons that isolate lay reasoning patterns from doctrinal recall or model sycophancy. It uses this methodology to successfully replicate with LLMs the results of human subject experiments.

Across three replications involving over 10,000 simulated judgments, leading models have internalized lay reasoning patterns that diverge from legal doctrine. Like human subjects, models prioritize social conformity over cost-benefit analysis when assessing negligence. They replicate the paradox that material deceptions, though judged more impactful to their receipients, are seen as *less* corrosive of consent than deceptions about a transaction's essential nature. And they track lay "contract formalism," perceiving hidden fees as more enforceable than fair, with judgments closer to ordinary consumers than to elite legal professionals.

Another finding is that models identify *what* tends to matter to ordinary people better than *how much* each factor matters. The numerical assessments that models give can amplify some effects and compress others. This and other limitations serve as the backbone for the recommendations the Article develops, alongside the call for future research and experimentation.

Understanding *why* these results hold is a question worth pausing on, particularly for those drawn to the psychology of human judgment. Humans make reasonableness judgments through deeply intuitive "system-1" processes, relying on mental schemas developed over a lifetime of socialization.[13] These frameworks are as hard to articulate as explaining how to tie shoelaces or hit a baseball.[14] While this makes them illegible to formal legal analysis and simple algorithms alike, they are, ironically, especially accessible to LLMs: industrial-grade pattern detectors trained on the same human discourse from which these schemas emerge.

The theoretical and practical applications of this research are broad.[15] Judges can check intuitions against simulated demographics before ruling on what "ordinary consumers" would understand. Regulators can pilot-test whether proposed disclosures communicate effectively, at a fraction of survey costs.[16] Resource-constrained litigants gain access to rough proxies for mock juries previously available only to wealthy parties. In each case, the tool functions not as final arbiter but as empirical reference point: a dictionary for reasonableness that makes departures from lay understanding transparent rather than hidden.

---

[11] See Koehn v. Delta Outsource Grp., Inc., 939 F.3d 863, 865 (7th Cir. 2019) ("One could say that, because Koehn did not read her employment contract, she never understood the terms and therefore was unable to assent to them."); *See also Solan et al., supra* note 4 (linking "ordinary meaning" claims to projection bias).

[12] All code and data used in this study are publicly available for replication, analysis, and follow-up work: https://github.com/yonathanarbel/generative-reasonable-person

[13] Linda Hamilton Krieger, *The Content of Our Categories: A Cognitive Bias Approach to Discrimination and Equal Employment Opportunity*, 47 STAN. L. REV. 1161, 1165, 1190 (1995) (introducing schema theory in legal contexts), building on David E. Rumelhart, *Schemata and the Cognitive System*, *in* 1 HANDBOOK OF SOCIAL COGNITION 71 (1984).

[14] *See e.g.,* Cheng, P.W. & Holyoak, K.J., *Pragmatic Reasoning Schemas*, 17 COGN. PSYCHOL. 391 (1985); Kevin J. Heller, *The Cognitive Psychology of Circumstantial Evidence*, 105 Mich. L. Rev. 241 (2006). My use of schema here is more narrow than its meaning in some of the social psychology literature; in my use, the focus is on latent mental models that are abstractive, but may be nonetheless complex (and even inconsistent), of even more complex social phenomena.

[15] *See* Calnan, *supra* note 9, at 82 ("Reasonable legal minds agree that reasonableness is one of the foundational concepts of American law, infiltrating everything from administrative, corporate, and constitutional law to crimes, torts, and contracts.").

The Article proceeds in four Parts. Part I explores the contested role and relevance of common reasonableness judgments in the adjudication of the reasonable person. It presents recent advances brought about by the experimental jurisprudence literature.

Part II locates this investigation within the burgeoning "silicon sampling" literature in economics, sociology, and psychology, which studies the ability of LLMs to replicate survey results.[17] Reviewing these developments, a recent Science article argued that "LLMs might supplant human participants for data collection."[18] Though not without limitations, social scientists find the method attractive because it offers unique advantages in terms of cost, scale, and reproducibility, and may even reduce certain forms of bias compared to traditional approaches.

Part III is the empirical heart of the Article. It begins with a methodological puzzle: how can we test whether LLMs have truly internalized lay reasoning rather than merely memorized legal doctrine? The problem is easy to overlook but difficult to solve. Models trained on vast legal corpora have encountered doctrinal definitions, case outcomes, and academic commentary. A model can recite the Hand formula or conclude that no reasonable person would view the Pepsi ad as serious, but this reveals nothing about whether it has absorbed the schemas that actually drive human judgments. It is the difference between a student who memorizes the answer key to an exam and one who learned the formulas.[19] Simply asking models to resolve cases or define reasonableness tests recall, not understanding.

The obvious remedy, presenting novel hypotheticals, introduces its own pathology. Models excel at detecting user intent and adjusting responses accordingly, a phenomenon researchers call "sycophancy." [20] If a prompt implies that two scenarios should be judged differently, the model may oblige regardless of its latent assessment. Even presenting scenarios side by side cannot be trusted: the model can harmonize its answers to appear consistent.[21] Probing latent understanding requires controlling for both memorization and demand effects

---

[16] See Igor Grossmann et al., *AI and the Transformation of Social Science Research*, 380 Science 1108 (2023) (arguing LLM-based approaches can reduce cost and frictions of data collection while changing the research pipeline); Marko Sarstedt et al., *Using Large Language Models to Generate Silicon Samples in Consumer and Marketing Research: Challenges, Opportunities, and Guidelines*, 41 *Psychol. & Mktg.* 21982 (2024) (discussing rapid, low-cost "silicon samples" as pretests, screeners, and iterative comprehension checks, with cautions)

[17] *See* Sarstedt et al., *supra* note 16, https://onlinelibrary.wiley.com/doi/10.1002/mar.21982.

[18] *See* Grossmann et al., *supra* note 16.

[19] *See* Randall Balestriero, Jérôme Pesenti & Yann LeCun, *Learning in High Dimension Always Amounts to Extrapolation*, ARXIV (2021), https://arxiv.org/abs/2110.09485; Timo Freiesleben & Thomas Grote, *Beyond Generalization: A Theory of Robustness in Machine Learning*, 202 SYNTHESE 109 (2023).

[20] OpenAI, Sycophancy in GPT-4O: What Happened and What We're Doing About It, https://openai.com/index/sycophancy-in-gpt-4o/ (Apr. 29, 2025)

[21] See OpenAI, *Sycophancy in GPT-4o: What Happened and What We're Doing About It* (Apr. 29, 2025) (discussing "sycophancy" and context-driven alignment with perceived user intent); Ricardo Dominguez-Olmedo, Moritz Hardt & Celestine Mendler-Dünner, *Questioning the Survey Responses of Large Language Models*, in *NeurIPS* 37 (2024) (showing sensitivity to question framing and prompt choices that can distort "survey-like" outputs); James Bisbee et al., *Synthetic Replacements for Human Survey Data? The Perils of Large Language Models*, *Political Analysis* (2024) (documenting instability and prompt dependence that undermines naïve survey substitution)

simultaneously. Yet the standard tools for doing so in human research, random assignment and between-subjects comparison, seem unavailable when the "subject" is a single model that remembers what it said moments ago.

The key is to exploit a feature of LLM architecture usually viewed as a limitation: *statelessness.* [22] When accessed through APIs rather than chat interfaces, each model session begins fresh with no memory of prior interactions. This transforms thousands of independent sessions into thousands of independent experimental subjects. Randomly assign conditions across sessions, and the model cannot harmonize, because no session knows what another session said. Present scenarios unlikely to appear in training data, and memorization cannot help. The result is Silicon Randomized Controlled Trials (S-RCTs): the methodological gold standard of social science,[23] adapted for AI. Part III develops this approach and applies it to three replications of human-subject studies spanning negligence, consent, and contract interpretation.

Part IV explores practical implications. Generative reasonable people do not replace judgment; they check it. Had Judge Wood consulted a silicon study in chambers, she might have reconsidered her confidence that teenagers would see the ad as mere jest. [24] She could still override the model on normative or prudential grounds, but the tool would surface what intuition obscured. In this sense, the LLM functions like a dictionary for reasonableness judgments, much as courts consult dictionaries to discern ordinary meaning. Regulators could pilot-test public understanding of disclosures at a fraction of survey costs. Resource-constrained litigants could access rough proxies for mock juries. Legal scholars could validate folk jurisprudence claims empirically. In each case, generative reasonable people function not as final arbiters but as empirical reference points: a starting place for inquiry rather than its conclusion.

This Article extends a research program exploring how generative AI can surface social facts relevant to adjudication. Prior work with David Hoffman examined whether models could illuminate patterns of language use in contract communities.[25] This Article asks a related but distinct question: whether models can approximate the reasonableness judgments—about negligence, consent, and contractual meaning—that legal standards invoke but rarely measure.

In the background of this Article, a larger conversation looms: should we let AI enter into the business of judging humans? Scholars have strong views on this question. [26] This Article will not resolve it, but it will clear the precondition for having it, clarifying the qualitative threshold that LLMs meet.

---

[22] For user applications, some AI providers now provide a memory feature, but this is not enabled in the current experiments.

[23] LAWRENCE M. FRIEDMAN, CURT D. FURBERG, AND DAVID L. DEMETS, FUNDAMENTALS OF CLINICAL TRIALS, at v. (4th ed, 2010)

[24] *See generally* Yonathan A. Arbel & Shmuel I. Becher, *Contracts in the Age of Smart Readers*, 90 GEO. WASH. L. REV. 83 (2022). (showing that large language models, working as "smart readers," can translate legal texts to be accessible to teenagers and translate among cultural divides.) *See also* Heinrich Peters & Sandra Matz, *Large Language Models Can Infer Psychological Dispositions of Social Media Users*, 3 PNAS Nexus e231 (2024), https://doi.org/10.1093/pnasnexus/pgae231.(finding that LLMs are especially capable of inferring psychological traits of younger individuals from their social media posts).

[25] On the use of LLMs as tools of interpretation, see Yonathan A. Arbel & David A. Hoffman, *Generative Interpretation*, 99 N.Y.U. L. REV. 451 (2024), recently discussed and adopted in *Snell v. United Specialty Insurance Co.*, 102 F.4th 1208 (11th Cir. 2024)

[26] *See* Kiel Brennan-Marquez and Stephen Henderson, *Artificial Intelligence and Role-Reversible Judgment*, 109 J. CRIM. L. & CRIMINOLOGY 137 (1019) (arguing that, independent of quality of decision, algorithms should not be in charge of adjudication based on—I believe—a quaint theory of role-reversibility as precondition to adjudication).

It also underscores the urgency of having it now, before practical compromises would force us. At the very least, the Article argues that, properly used, LLMs offer a sufficiently reliable improvement over current methods to justify careful integration into legal practice.

## I. FOLK OPINIONS AND THE LAW

How much should the law account for lay perceptions of reasonableness? The law has a basic duality. Judged by its mode of production, the law is clearly a formal, technocratic, and in some sense elitist enterprise.[27] It relies on a cadre of professionals—judges, legislators, regulators—to mediate its operation. It follows specific rules that help shape its meaning, internal coherence, administration, and effectiveness. All of this involves jargon, terms of art, and specialized language: ***noscitur a sociis, habeas corpus, proximate cause.*** On occasion, the law also coopts common terms such as *contract, tangible,*[28] or *fish.*[29]

Yet to reduce law to the sausage-making formalities of its production would distract from its practical ability to govern individuals that neither speak Latin nor particularly care about Hand's formula. That is, the law is also a social phenomenon, and it aims to speak in the language of the governed.[30] It is this social aspect that motivates many scholarly and reform proposals that push against the specialized language of the law.

Consider a few scholarly conversations that tap ordinary opinions. The technocratic plain language movement, the largest consumer reform movement of our generation, sought to rewrite the language of the law to match the common vernacular.[31] A more scholarly enterprise, the folk jurisprudence project, seeks to map the lay understanding of legal concepts and measure the divergence of lay and lawyerly understanding.[32] Likewise, a central method of legal interpretation, "ordinary meaning" analysis, hews closely to lay usage of language.[33] Trends in criminal justice, perhaps most perniciously penal populism, seek to adjust sanctions to folk sense of desert and punishment.[34] In contrast, the recent "lived experience" scholarship attempts to surface lay experiences of marginalized people into the study of the law.[35]

The commitment to lay perceptions transcends the pragmatic and informs jurisprudence itself. For H.L.A Hart, lay intuitions form jurisprudence's essence: "a general theory of law is just an attempt to elucidate the folk concept of law."[36] Joseph Raz would later trace this lineage, arguing jurisprudence seeks "our ordinary concept of law"—not as scholars define it, but as bus drivers and IT professionals live it.[37] Other scholars may espouse a more elitist or technocratic view of the law, but almost all agree that the law should be mindful of, if not always reducible to, lay attitudes.[38]

---

[27] This is, in essence, the Hartian view of legal norms under the rule of recognition. For a fuller treatment, see Felipe Jimenez, *Legal Principles, Law, and Tradition*, 33 YALE J. L. & HUMAN. 59 (2020).

[28] C.R.S.A § 39-26-102 (2022) 15(b.5(I) (""Tangible personal property" includes digital goods").

[29] *See* Yates v. United States, 574 U.S. 528 (2015) (discussing whether fish is a "tangible good" for purposes of section 1519 of the Sarbanes-Oxley Act of 2002).

[30] THE DECLARATION OF INDEPENDENCE para. 2 (U.S. 1776) ("[T]o secure these rights, Governments are instituted among Men, deriving their just powers from the consent of the governed"). In an insightful article, Anya Bernstein argues against a narrow language of the governed view which she positions within an Austinian language-as-command jurisprudence, and suggests that the audiences of legal language are often government agencies. Anya Bernstein, *Legal Corpus Linguistics and the Half-Empirical Attitude*, 106 CORNELL L. REV. 1397, 1435 (2021).

[31] *See* Yonathan A. Arbel, *The Readability of Contracts: Big Data Analysis*, SSRN.

[32] *See generally* Kevin Tobia, *Experimental Jurisprudence*, 89 U. CHI. L. REV. 735 (2022)

[33] *See e.g.,* Oliver Wendell Holmes, *The Theory of Legal Interpretation*, 12 HARV. L. REV. 417, 417 (1899)("[W]e ask, not what this man meant, but what those words would mean in the mouth of a normal speaker of English…"); Richard S. Kay, *Original Intention and Public Meaning in Constitutional Interpretation*, 103 NW. U. L. REV. 703, 719 (2009) ("By definition, the public meaning of a rule is the one apparent to a competent speaker of the language from a mere inspection of the text.").

[34] *See* Jocelyn Simonson, *Police Reform Through a Power Lens*, 130 YALE L.J. 778, 850 (2021).

[35] *See* Rachel López, *Participatory Law Scholarship*, 123 COLUM. L. REV. 1795 (2023).

[36] *See* Brian Leiter & Alex Langlinais, *The Methodology of Legal Philosophy*, *in* THE OXFORD HANDBOOK OF PHILOSOPHICAL METHODOLOGY 467 (Herman Cappelen, Tamar Gendler & John Hawthorne eds., Oxford Univ. Press 2016).

The evolution of the reasonable person concept in tort is illuminating. Making its debut appearance in 1837, the reasonable person was first conceived of as a "a man of ordinary prudence."[39] According to Rabin, that idea of fault in tort law was originally tied "to community expectations of reasonable behavior, rather than to the economist's perception of rational behavior." [40] That is, negligence standards were originally construed as anchored in empirical facts, in particular, exogenously determined community norms.[41] It will not be until the Twentieth century that judges like Learned Hand would take a decidedly normative approach:

> [I]n most cases reasonable prudence is in fact common prudence; but strictly it is never its measure. ... Courts must in the end say what is required; there are precautions so imperative that even their universal disregard will not excuse their omission.[42]

Today, these debates are still ongoing, with some taking a descriptive view, others prescriptive, and yet others, some hybrid of the two.[43] But regardless of jurisprudential commitments, the concept of reasonableness is never more than one degree of separation from lay opinions.[44] This is for a combination of descriptive, pragmatic, and normative reasons. The first reason is reflective: when we know what lay people truly think, we gain a better understanding of what legal concepts mean.[45] The second is effectiveness: if the law sets to direct behavior, it should speak in the language of the governed, or at least be attuned to how it is being heard. This is part of the animus of the ordinary meaning interpretive theory. [46] The third is legitimacy: for people to trust the law, they should be able to understand it.[47] This is closely related to participatory arguments, about the public's role in shaping their lives. The fourth is political. If the public is to set a check on the operation of the legal system, it is important that it understand its laws, commands, and boundaries.[48]

---

[37] *See* JOSEPH RAZ, PRACTICAL REASON AND NORMS 164 (2d ed. 1999). *Cited in Jimenez, supra* note 9

[38] *See Jimenez, supra* note 9.

[39] Vaughn v. Menlove (1837) 132 ER 490 (CP).

[40] *See* Robert L. Rabin, *The Historical Development of the Fault Principle: A Reinterpretation*, 15 GA . L. REV. 925, 931 (1981)

[41] For an articulation of the positivist view, *see* Alan D. Miller & Ronen Perry, *The Reasonable Person*, 87 N.Y.U. L. REV. 323, 370-2 (2012).

[42] The T.J. Hooper, 60 F.2d 737, 740 (1932) at 740.

[43] Lucien Baumgartner & Markus Kneer, The Meaning of "Reasonable": Evidence From a Corpus-Linguistic Study 1–2 (Aug. 29, 2023), https://ssrn.com/abstract=4555547; Benjamin C. Zipursky, Reasonableness in and out of Negligence Law, 163 U. Pa. L. Rev. 2131, 2150 (2015) (proposing a hybrid view).

[44] *See also Baumgartner & Kneer, supra* note 43, at 4 ("there are grounds for (considerable) correspondence between the lay concept of reasonableness and its legal equivalent").

[45] *See Tobia, supra* note 32, at 750 (Presenting "the "folk-law thesis." . . . this account would predict that the legal concept of causation reflects features of the ordinary concept of causation and that the legal concept of consent reflects features of the ordinary concept of consent.").

[46] *See Tobia et al., supra* note 6, at 372 (noting the motivations behind modern textualism include "concern for democracy, fair notice, or rule of law values, or objective inquiry into meaning").

[47] TOM TYLER, WHY DO PEOPLE OBEY THE LAW (2006) at 7 (if people "regard legal authorities as more legitimate, they are less likely to break any laws . . . A normative perspective leads to a focus on people's internalized norms of justice and obligation. It suggests the need to explore what citizens think and to understand their values")

[48] Jason M. Solomon, *The Political Puzzle of the Civil L Jury*, 61 EMORY L.J. 1331, 1340 (2012) ("Historically, the civil jury in the United States, like the criminal jury, was justified in large part as a check against the abuse of government power."

There is also the more Hayekian reason.[49] Disperse individuals have access to information not available to the central planner. Lived experience and peer-to-peer interactions produce perspectives and knowledge that are not legible to either well-meaning policymakers or well-read scholars. Aggregating this information leverages the wisdom of the crowds, potentially creating judgments more accurate than that of any specific individual.[50]

Thus, it may be the case that the more modern tort view is that the reasonable person is essentially normative creation, a man fashioned in the judge's image.[51] Yet, even if we are all realists now, and even if we all understand now the reasonable person as a normative construct, it is still incumbent upon on us to reflect on how that judicial person relates to actual people. The descriptive need not determine the normative; but it can inform it.[52]

But this conclusion presents the challenge: how can the State make ordinary opinions legible to itself?[53] How can the law discover what lay people think? And in a democracy, where the people rule, how do *we* know what people think?

---

[49] Friedrich A. Hayek, *The Use of Knowledge in Society*, 35 AM. ECON. REV. 519 (1945) ("The knowledge of the circumstances of which we must make use never exists in concentrated or integrated form, but solely as the dispersed bits of incomplete and frequently contradictory knowledge which all the separate individuals possess.").

[50] A recent prediction competition is illustrative of this phenomenon. Asked to predict multiple future events, the average participant (N=3,300) ranked slightly worse than chance. The average aggregate prediction, however, ranked at the 95th percentile of all participants. https://www.astralcodexten.com/p/who-predicted-2023.

[51] *See e.g.*, Mayo Moran, *The Reasonable Person: A Conceptual Biography in Comparative Perspective*, 14 LEWIS & CLARK L. REV. 1234, 1236 (2010) ("both in the context of the law of negligence and in the criminal context, the objective content of the reasonable person is closely linked to standards of ordinariness or normalcy").

[52] ee Brian Z. Tamanaha, *Beyond the Formalist-Realist Divide: The Role of Politics in Judging* (2010) (arguing realism's insights have become part of mainstream judging debates); Brian Leiter, *Legal Realism and Legal Positivism Reconsidered*, 111 *Ethics* 278 (2001) (describing realism's enduring influence on contemporary legal theory)

[53] On legibility as a central goal of the state, see JAMES C. SCOTT, SEEING LIKE A STATE (1998)

### 2. GENERATIVE PEOPLE IN THEORY AND THE SOCIAL SCIENCES

LLMs are, at core, industrial-grade pattern recognizers.[54] The idea behind generative reasonable people is that, because of this, they might also pick up on the subtle, complex, and perhaps even self-contradictory patterns that drive the everyday judgments of individuals on matters of reasonableness.[55]

This claim may seem ambitious at first. Legal scholars are already familiar with the effectiveness of statistical models to predict case outcomes.[56] For instance, even though Supreme Court cases are involved and complex, a single factor—political affiliation of the Justice's nominating President—will be very predictive of outcomes in political cases.[57] The literature on structural bias also shows that race and socioeconomic status also hold predictive power, even if they really should not.[58] Similarly, machine learning models excel at predicting consumer behavior in advertising and fraud detection.[59] Yet, predicting reasonableness judgments, which appear so closely related to *ethical* and *social* judgments, may seem qualitatively different. Why should we expect generative AI models to achieve even modest predictive accuracy on nuanced questions of reasonableness?

This Part offers the theoretical and empirical background necessary to assess this possibility. The first section explores how modern LLMs' architecture and emergent capabilities, such as attention mechanisms, roleplaying abilities, and generalization, make them potentially suitable for simulating ordinary people's judgments. These same architectural features, however, also introduce systematic limitations: a majoritarian bent that may marginalize minority perspectives, susceptibility to bias amplification, and "value drift" as social norms evolve beyond the models' training data. The second section reviews empirical evidence from the burgeoning "silicon sampling" literature in psychology, economics, and sociology, which demonstrates that LLMs can replicate human survey responses across diverse domains with surprising fidelity. Taken together, these developments suggest cautious optimism: generative reasonable people are theoretically plausible and empirically promising, but their deployment demands careful attention to their inherent limitations. As to whether generative reasonable people work in practice is addressed empirically in the next Part.60

---

[54] For more on the distinction between simulation and prediction, see Yonathan A. Arbel, *Time & Contract Interpretation: Lessons from Machine Learning*, *in* RESEARCH HANDBOOK ON LAW AND TIME (Frank Fagan & Saul Levmore eds., forthcoming 2024).

[55] This fallacy, that AI analysis constitutes a form of judgment rather than pattern recognition, has led to some confused commentary that this Article decidedly avoids. *See* Brennan-Marquez & Henderson, *Artificial Intelligence and Role-Reversible Judgment*, 109 J. CRIM. L. & CRIMINOLOGY ART. 1 (2019).

[56] *See, e.g.,* Theodore W. Ruger et al.,*The Supreme Court Forecasting Project: Legal and Political Science Approaches to predicting Supreme Court Decision Making*, 104 COLUM. L. REV., 1150–1210 (2004) (predicting, with 76% accuracy, case outcomes based on sparse factors). Kimo Gandall, Chris Haley, Juliana Chhouk, Logan Knight, Alex Wang, and Bella DeMarco, *Predicting Precedent: A Psycholinguistic Artificial Intelligence in the Supreme Court*, 14 220 (2023) (offering a modest improvement, but at the cost of a complex model).

[57] *See* Jeffrey A. Segal & Alan J. Champlin, *The Attitudinal Model*, *in* ROUTLEDGE HANDBOOK OF JUDICIAL BEHAVIOR 29 (2017) ("The attitudinal model is the most dominant model for understanding the Supreme Court's decisions on the merits. In fact, for the eight justices currently on the Court prior to the 2016 term, the correlation between their ideology and their voting behavior on the Court is a .94").

[58] U.S. Sentencing Comm'n, *Demographic Differences in Federal Sentencing: An Update to the 2012 Booker Report* (2023) (finding Black male offenders receive on average 13.4% longer sentences than comparable White males); Will Dobbie, Jacob Goldin & Crystal S. Yang, *The Effects of Pretrial Detention on Conviction, Future Crime, and Employment*, 108 AM. ECON. REV. 201, 203–04 (2018) (causal evidence that inability to pay bail increases conviction probability and worsens outcomes).

[59] *See* John Ford et al., *AI Advertising: An Overview and Guidelines,* J. BUS. RSCH. 166 (2023).

### 1. Generative AI and Silicon Jurors: Emergent Capabilities

The foundation of generative reasonable people lies in the sophisticated architecture of modern artificial intelligence systems, particularly in the realm of natural language processing.[61] Four key capabilities make them suitable for the task of modeling reasonableness judgements: attention mechanisms, emergent roleplaying capabilities, generalization abilities, and their "majoritarian bent".[62]

Current generative AI architectures, including those fit for generative reasonable people, rely on autoregressive models.[63] These models generate output sequentially, with each token (roughly, a word) conditioned on those previously generated.[64] During training, the models are fed large volumes of data—more text than any human can read in a lifetime—and are tasked with predicting the next token in a sequence of words. The model learns by minimizing prediction errors, gradually improving its ability to anticipate what comes next in human language.

The true breakthrough that catapulted language models to their current capabilities was the introduction of the transformer architecture.[65] At its heart is the attention mechanism, which allows the model to dynamically weigh the importance of different parts of the input.[66] In this architecture, the values assigned to the vector representation of each token are adjusted based on contextual relationships. A token like "sea" would be described by various numbers, indicating its relationship to concepts like water, ships, and Poseidon. These numbers adjust based on context: desalination, circumnavigation, or Odysseus' travails.

The importance of this becomes apparent when we consider the word 'bank.' On its own, it is ambiguous: a financial institution or the side of a river? But when humans see a sentence like "Frank needed money so he went to the bank", they immediately adjust their understanding of the word based on context. So does the model; the attention mechanism shifts the meaning of "bank" towards financial institution when it encounters the word 'money' in this sentence. Just as a judge might focus on key elements of a case while considering the broader context, AI attention mechanisms allow models to prioritize relevant information when making judgments - crucial for assessing reasonableness in complex legal scenarios.

One of the most fascinating aspects of these models is their *emergent* capabilities. An emergent property is one that appears only at a given level of complexity.[67] The ability to roleplay character is an emergent property and modern models perform well on this task.[68] No programmer explicitly coded rules for this behavior - rather, roleplaying emerges naturally from the system's fundamental function of next-token prediction. If a sentence mentions that the speaker is Chris Tarrant, the probability shifts toward predicting "Is that your final answer" as the next phrase. Context affects prediction, and identity informs context.[69]

---

60 *See infra* Part III.

[61] At this point, there is no dearth of introductory materials at different levels of technical expertise. For an overview of the rapid improvements in the field, see Aji Supriyono, Aji Prasetya Wibawa, Suyono & Fachrul Kurniawan, *Advancements in Natural Language Processing: Implications, Challenges, and Future Directions*, 16 TELEMAT. & INFORMAT. REP. 100173 (2024)

[62] Yonathan A. Arbel & David A. Hoffman, Generative Interpretation, 99 N.Y.U. L. Rev. 451, 476–83 (2024).

[63] For an introduction geared towards lawyers, *see Arbel & Hoffman, supra* note 62, at 476–83.

[64] tokens are commonly appearing word subparts, such as 'th' in English . A helpful list of all the 100,00 tokens used by GPT-4 can be found here: https://gist.github.com/s-macke/ae83f6afb89794350f8d9a1ad8a09193.

[65] *See* Ashish Vaswani et al., *Attention Is All You Need*, arXiv:1706.03762 (2017).

[66] The earlier models did not use attention mechanisms, but given the dominance of transformers today, I focus on them.

Roleplaying is crucial for generative reasonable people. This capability allows models to produce responses that cohere with broad patterns of reasoning among common people and shift from "expert" mode to layperson mode. It instructs the model to move from its default, "helpful assistant" voice, to more realistic depictions of ordinary people.[70] This transition is essential if we seek a non-elitist notion of reasonableness, as models have internalized both expert and lay patterns of reasoning.

Generalization is another crucial emergent property of these systems. A model like LLaMA-3 has 70 billion parameters but it is trained on 15 trillion tokens.[71] The ratio is one parameter for every 214 tokens. This means that rote memorization of all the data the model sees during training is impossible. Instead, the model must learn to compress the information by generalizing the patterns it sees. This is similar to how humans learn abstract rules rather than memorizing the details of every specific instance. We might not remember each cat we have seen, but we learn to identify them by generalizing the concept of a cat from its specific instances. In other words, we develop a model of "catness," and while we will be hard-pressed to articulate it, it allows us to quickly and efficiently identify cats even in novel situations.[72]

Generalization is vital for generative reasonable people because many questions will involve scenarios different from those in the training data. The hope is that models have generalized ideas about reasonableness rather than simply memorizing specific instances when an act was deemed reasonable or unreasonable.[73] While generalization is necessary, it doesn't guarantee success - it may be superficial, crude, or mistaken. What matters is that AI models can develop complex internal models beyond simple pattern recognition.

Even though generalization may be necessary to the task at hand, it doesn't guarantee success. Generalization may also be superficial (overfitting), crude (underfitting), or simply mistaken. This means that we would want to test both the existence of a generalized model *and* its adequacy. But the key for now is to understand that AI models can learn more than simple patterns in data, and they can develop internal models that are more complex. (in fact, many complaints about algorithmic black boxes show that these internal models may be too complex).

---

[67] *See* Yonathan A. Arbel, *Reputation Failure: The Limits of Market Discipline in Consumer Markets*, 54 WAKE FOREST L. REV. 1239, 1252 n. 64 (2019).

[68] *See* Zekun Moore Wang et al., *RoleLLM: Benchmarking, Eliciting, and Enhancing Role-Playing Abilities of Large Language Models*, ARXIV preprint (2024), https://arxiv.org/abs/2310.00746 ("State-of-the-art (SOTA) LLMs like GPT-4 . . . exhibit advanced role-playing capabilities"); Keming Lu et al., *Large Language Models are Superpositions of All Characters: Attaining Arbitrary Role-play via Self-Alignment*, arXiv preprint (2024), https://arxiv.org/abs/2401.12474 ("GPT-4 has already demonstrated outstanding role-playing abilities"); Jiangjie Chen et al., *From Persona to Personalization: A Survey on Role-Playing Language Agents*, ARXIV preprint (2024), https://arxiv.org/abs/2404.18231 ("Personas are inherent in LLMs, and role-playing them capitalizes on the statistical stereotypes in LLMs").

[69] Another technical aspect that contributes to the success of roleplaying activities is instruction-tuning of models, which improves their ability to stay in character. *See Chen et al., supra* note 68.

[70] *See Lu et al., supra* note 68 (positing that roleplaying arises from a combination of training data and in-context learning).

[71] *See* Meta, *Introducing Meta Llama 3: The Most Capable Openly Available LLM to Date*, Meta AI (Apr. 18, 2024), https://ai.meta.com/blog/meta-llama-3/.

[72] Reddit, "r/CatsInWeirdPlaces," https://www.reddit.com/r/CatsInWeirdPlaces/.

[73] The phenomenon of generalization is also known as "grokking" and the study of the points in training where models "grok" new concepts is an active area of research. See e.g., Hu Qiye, Zhou Hao & Yu RuoXi, *Exploring Grokking: Experimental and Mechanistic Investigations*, ARXIV:2412.10898 (2024), https://arxiv.org/abs/2412.10898.

A final intriguing characteristic is what we might term their "majoritarian bent." [74] Models favor broader patterns over narrower ones, manifesting as a pro-majority bias. This arises from two factors: the statistical nature of next-token prediction inherently favors common patterns, and post-training adaptations like Reinforcement Learning from Human Feedback further align models with general human preferences. This majoritarian tendency makes these models well-suited for simulating the "reasonable person" standard, as they naturally gravitate toward common opinions and widely-held beliefs.75

While the majoritarian bent is essential to the utility of generative reasonable people, it also points to the limits of this technique. The models mirror aggregated human knowledge and biases, including problematic ones. [76] In legal contexts, they may struggle with cases requiring consideration of diverse or minority perspectives - a concern highlighted by critical scholars. Feminist legal theorists have exposed how the supposedly neutral "reasonable person" has often been the "reasonable man" in practice, with majoritarian defaults imposing asymmetric standards that present majority experiences as the natural baseline.[77] Generative reasonable people may well replicate such patterns. [78]

Two other limitations deserve emphasis. First, general models struggle to reliably simulate specific individuals - they face a "granularity problem."[79] While some applications attempt to emulate specific people through fine-tuning and context learning, [80] the effectiveness of such approaches remains unproven. Second, pretrained models have a limited "half-life" as social norms and perceptions change, creating "value drift" that makes older models incapable of reflecting societal evolution. [81]

To summarize, there is a surprising but deep connection between modern AI architecture and generative reasonable people. These models can understand context, adopt various perspectives, generalize from examples, and reflect common social norms. However, they also have notable limitations regarding bias, individuality, and temporal relevance - considerations that must inform their application to legal questions.

---

[74] *See Arbel & Hoffman, supra* note 62.

75 *Id.*

[76] *See generally* Sandra G. Mayson, *Bias In, Bias Out*, 128 YALE L.J. 2122 (2019).

[77] *See* Susan Dimock, *Reasonable Women in the Law*, 11.2 CRIT. REV. INT'L SOC. & POL. PHIL. 153, 153 (2008) ("What counts as reasonable in these and many other areas of law is typically conceptualized against a 'reasonable man' . . . standard"). Dimock argues that even lower-level abstractions, like 'reasonable woman', are still over-generalized.

[78] Psychometric analysis suggests that "LLMs exhibit a tendency toward Undifferentiated, with a slight inclination toward Masculinity." Jen-tse Huang et al., *On the Humanity of Conversational AI: Evaluating the Psychological Portrayal of LLMs*, *in* Proc. of the Int'l Conf. on Learning Representations (ICLR) (2024).

[79] *See Wang et al., supra* note 68.

[80] For an excellent review of the literature studying the differences between parametric and non-parametric roleplaying, *See Chen et al., supra* note 68, https://arxiv.org/abs/2404.18231.

[81] A feature and a bug: having time frozen models can also be useful in depicting the attitudes of older generations. Perhaps some arguments about originalism could have been resolved had we had a powerful model trained on materials from that time.

### 2. Silicon Sampling in Social Sciences

A growing body of research on "silicon sampling" demonstrates LLMs' ability to provide human-like responses across various domains of social science. It has led to some interesting discoveries on the power of AI to provide human-like feedback in various areas of the social sciences and adds to the plausibility of generative reasonable people. [82]

The evidence is striking. One study found that LLMs can generate moral judgments highly correlated (r=0.95) with human judgments.[83] Another showed that on eleven sociological questions, LLM responses closely aligned with those of the general population.[84] Early versions of ChatGPT successfully replicated multiple psychological studies.[85] Another psychological study found that LLMs can persuasively assume big five personality traits such as extroversion or agreeableness.[86] perhaps most impressively, a recent study found that "ChatGPT-4 exhibits behavioral and personality traits that are statistically indistinguishable from a random human from tens of thousands of human subjects from more than 50 countries."[87]

LLMs even replicate human cognitive biases, bringing them closer to actual human judgment.[88] For example, one study found that LLMs can recreate classic findings in economics and psychology, such as the ultimatum game and the Milgram obedience experiment.[89] Interestingly, models sometimes display more ethical behavior than humans - showing less selfishness and greater fairness toward out-group members, raising questions about whether we want perfect mimesis or idealized behavior.[90]

The roleplaying capabilities of LLMs offer particularly promising applications for legal analysis. Models can be prompted to respond as reasonable persons from various demographic backgrounds or even to simulate the reasoning of historical legal figures. This capability has proven so compelling that companies like Character.ai have built billion-dollar companies around it, offering interactive experiences with simulated personas.[91]

---

[82] *See e.g., See Sarstedt et al., supra* note 16.

[83] *See* Danica Dillion, Niket Tandon, Yuling Gu & Kurt Gray, *Can AI Language Models Replace Human Participants?* 28 TRENDS COGN. SCI. 597 (2023). In hindsight, it is not entirely surprising because these methods are trained to mimic human moral judgments using RLHF and similar techniques.

[84] *See* James Bisbee et al., *Synthetic Replacements for Human Survey Data? The Perils of Large Language Models*, *in* Political Analysis (Published online 2024:1-16.) doi:10.1017/pan.2024.5 The researchers find low accuracy regarding the distribution of synthetic opinions, a point we revisit later.

[85] *See* Peter S. Park et al. *Diminished diversity-of-thought in a standard large language model*, arXiv:2207.07051 (2023).

[86] *See* Hang Jiang, Xiajie Zhang, Xubo Cao, Cynthia Breazeal, & Jad Kabbara, *PersonaLLM: Investigating the Ability of Large Language Models to Express Big Five Personality Traits*, arXiv:2305.02547v5 (2023), https://doi.org/10.48550/arXiv.2305.02547.

[87] Qiaozhu Mei, Yutong Xie, Walter Yuan, & Matthew O. Jackson, *A Turing Test of Whether AI Chatbots Are Behaviorally Similar to Humans*, [Small Caps]Proc. Nat'l Acad. Sci. U.S.A.[/Small Caps], Feb. 22, 2024, at e2313925121, https://doi.org/10.1073/pnas.2313925121.

[88] See Andrew K. Lampinen et al., Language Models Show Human-Like Content Effects on Reasoning Tasks, arXiv:2207.07051 (2022), https://doi.org/10.48550/arXiv.2207.07051 (showing that the framing of questions misleads humans and LLMs in similar ways); *See also Park et al., supra* note 85 (showing false consensus bias).

[89] Gati Aher, Rosa I. Arriaga & Adam T. Kalai, "Using Large Language Models to Simulate Multiple Humans and Replicate Human Subject Studies," in *Proc. of the 40th Int'l Conf. on Machine Learning* (ICML 2023)

[90] *See Huang et al., supra* note 78 (finding that "LLMs demonstrate reduced ICB scores compared to the general human population." The ICB scale is a measure of an "individual's belief in whether their ethnic culture predominantly shapes a person's identity").

But how accurate are these simulations? In one validation study using a "personal Turing test," AI models imitating specific individuals achieved a 48.3% success rate in deceiving acquaintances of those individuals.[92] This is quite remarkable: in half of the cases, acquaintances could not tell apart a model from the actual person.[93]

Research shows these capabilities can be enhanced through various techniques.[94] Giving personas demographically typical names improves performance.[95] One study prompted LLMs to assume the persona of people with specific demographic characteristics and answer a few questions. Then they asked humans to answer the same questions, some of them met these demographics (in-group), some of them assumed the persona of that person (out-group). The researchers find that in some instances, LLMs sound more like out-group than in-group members.[96] This capability is improved if the model is given a name that is consistent with the underlying demographic.[97] Further, injecting randomness into responses helps prevent "group flattening" - the tendency to produce stereotypical answers for minority groups.[98]

This is consistent with another recent research paper that found that in situations where humans have polarized views, persona assignment helps the model express differing views.[99] Finally, and in the other direction, a recent study suggests that many results that align with specific demographic merely reflect prompting effects and disappear when prompts are carefully vetted.[100]

While the focus here is on the generative reasonable person, the research also points toward the possibility of generative juries. Studies show that when LLM agents interact, they exhibit group dynamics similar to human collectives, modeling complex social phenomena like bank runs,[101] realistic macroeconomic phenomena,[102] information cascades,[103] and community formation.[104]

---

[91] *Character.AI In Early Talks for Funding at More Than $5 Billion Valuation*, Bloomberg (September 28, 2023, 4:43 PM CDT) https://www.bloomberg.com/news/articles/2023-09-28/character-ai-in-early-talks-for-funding-at-more-than-5-billion-valuation

[92] *See* Man Tik Ng et al., *How Well can LLMs Echo Us? Evaluating AI Chatbots' Role-Play Ability with ECHO*, arXiv:2404.13957 [cs.CL] at 7 (Apr. 22, 2024), https://arxiv.org/abs/2404.13957.

[93] See Man Tik Ng et al., *How Well Can LLMs Echo Us? Evaluating AI Chatbots' Role-Play Ability with ECHO* (2024) (reporting "personal Turing test" results measuring whether acquaintances can distinguish chatbot role-play from the person); Qiaozhu Mei et al., *A Turing Test of Whether AI Chatbots Are Behaviorally Similar to Humans*, 121(9) *Proc. Nat'l Acad. Sci.* e2313925121 (2024) (cross-national behavioral similarity tests and distributions).

[94] *See e.g.,* Cheng Li et al., RoleLLM: Benchmarking, Eliciting, and Enhancing Role-Playing in Large Language Models, arXiv:2310.00746 (2023), https://arxiv.org/abs/2310.00746 Alireza Salemi et al., LaMP: When Large Language Models Meet Personalization, arXiv:2304.11406 (2023), https://arxiv.org/abs/2304.11406; Nilimesh Halder, *Harnessing the Power of Role-Playing in Advanced AI Language Models: A Comprehensive Guide to ChatGPT's Potential*, Medium (2023),

[95] Angelina Wang, Jamie Morgenstern & John P. Dickerson, *Artificial Intelligence Chatbots Mimic Human Collective Behaviour*, arXiv:2402.01908v1 [cs.CY] (Feb. 2, 2024), https://arxiv.org/abs/2402.01908v1.

[96] *Id.*

[97] *Id* at figure 3.

[98] *Id.*

[99] Tiancheng Hu & Nigel Collier, *Quantifying the Persona Effect in LLM Simulations*, arXiv:2402.10811 (Feb. 26, 2024), https://arxiv.org/abs/2402.10811.

[100] Ricardo Dominguez-Olmedo, Moritz Hardt & Celestine Mendler-Dünner, *Questioning the Survey Responses of Large Language Models*, *in* Advances in Neural Information Processing Systems 37 (2024).

Of course, limitations remain.[105] Quality roleplaying requires sufficient contextual information about the individuals or groups being simulated. Current models struggle with some types of reasoning, particularly around politically charged topics. Prompting models to act as a typical member of a group risks reinforcing simplistic or stereotypical portrayals of complex groups. And ethical concerns about "speaking for" practices must be addressed,[106] particularly when simulating marginalized groups.

Despite these challenges, the silicon sampling literature provides substantial evidence that LLMs can approximate human-like judgments across diverse domains. This capability forms the foundation for our exploration of generative reasonable people in legal contexts.

---

[101] Sophia Kazinnika, *Bank Run, Interrupted: Modeling Deposit Withdrawals with Generative AI*, Federal Reserve Bank - Quantitative Supervision & Research (Oct. 30, 2023), https://papers.ssrn.com/sol3/papers.cfm?abstract_id=4656722.

[102] Ningyuan Li, Chong Gao, Yiming Li, & Qi Liao, *Large Language Model-Empowered Agents for Simulating Macroeconomic Activities*, arXiv:2310.10436 (2023).

[103] *See Huang et al., supra* note 78.

[104] James He et al., *Artificial Intelligence Chatbots Mimic Human Collective Behaviour*, preprint at Research Square (2024).

[105] *See supra* note 86 and accompanying text.

[106] Linda Martín Alcoff, *The Problem of Speaking for Others*, CULTURAL CRITIQUE, No. 20, 5–32 (1991).

## 3. THE REASONABLE PERSON IN SILICON

This Part offers an empirical assessment of generative reasonable people. As noted in the Introduction, evaluating models' deeper latent understanding of concepts runs the risk of learning-to-the-test, contamination, and sycophancy.[107] To address these issues, the methodology developed, S-RCT, exploits the stateless nature of large language models and uses them to adapt the standard method of randomized controlled trials. By testing models under controlled conditions mirroring human studies, we can probe their latent reasoning, not just their ability to parrot precedent.

### 1. The Core Methodology: S-RCT

The foundational challenge confronting any attempt to use AI as an instrument of applied jurisprudence is methodological: How can we know what LLMs truly understand about reasonableness rather than what they have memorized?

Two distinct threats to validity confront naïve attempts to answer this question. The first is "learning to the test" or "contamination". Models trained on legal corpora may have encountered doctrinal definitions, case outcomes, or even the academic studies we seek to replicate. Asking a model to define reasonableness or resolve *Leonard v. Pepsico* is a test of recall, not generalization. The models are very likely to succeed at such tasks, but this will not answer the question.

The second threat is demand effects or "sycophancy." To solve the previous issue, we might present models with unseen hypotheticals and than assess their responses. The problem is that models excel at inferring user intent and adjusting responses accordingly. Thus, the models might give the right answer not because it learned that a factor is important but because the question structure might imply that. Even putting side-by-side scenarios will not resolve the issue. If a model sees two scenarios side by side, it may infer that we expect differential treatment and harmonize its answers to appear consistent. To probe a model's true latent representations, we need methods that control both threats simultaneously.

The solution emerges from an architectural feature of modern LLMs typically viewed as a limitation: their fundamental statelessness. When interacting with LLMs through APIs rather than standard chat interfaces, each conversation starts from a blank slate with no memory of prior interactions. This creates an unexpected experimental opportunity. We can treat each independent model session as a distinct experimental subject, enabling adaptation of the gold standard of causal inference to the study of AI reasoning: silicon randomized controlled trials (S-RCT).

S-RCT rests on three mechanisms. The first is statistical independence through statelessness. Thousands of separate model sessions receive random assignment to experimental conditions with no ability to communicate or coordinate responses. A session assigned to one condition cannot observe, and therefore cannot harmonize with, sessions assigned to other conditions.

[107] *See supra* Introduction.

The second mechanism is the shift from absolute to differential measurement. Rather than asking models to produce judgments vulnerable to memorization and calibration problems, we measure how responses change between randomly assigned conditions. Even if a model has learned from case law that "reasonable care requires salting icy sidewalks," it cannot anticipate our specific experimental manipulation or know what score another session produced under different conditions. The differential approach isolates the model's sensitivity to manipulated factors from any memorized baseline.

We say a model is *aligned* to human judgment if it shifts its judgments in the same direction humans do when experimental parameters vary. This sidesteps the problem of comparing raw scores across different types of reasoners. We need not ask whether a model's "6 out of 10" means the same as a human's. We ask only whether both shift upward when a precaution becomes more common, and whether that shift is statistically reliable.

The third mechanism is persona assignment. Models are instructed to roleplay demographically specified individuals, injecting realism and variation into the decision process. For robustness, the studies here test whether personas improve alignment with human baselines and verify results across multiple models: GPT-5 Chat, DeepSeek R1, Grok-4-Fast, Claude Sonnet 4.5, Gemini 2.5 Pro, Kimi K2, Qwen 3 max, and Mistral Medium. This selection spans open-source and proprietary architectures, different training approaches, and varying parameter counts, offering an overview of today's state of the art.

**Sample Persona Instructions**

You are roleplaying as Mary Alvarez. Alvarez is a 61 year old Hispanic woman. Politically, she is Lean Democrat. Personality description: Mary Alvarez, at 61 with a high school education, exhibits a down-to-earth yet pragmatic persona. Her modest yet comfortable lifestyle as an employed individual within the $75K-$150K income bracket reflects her practicality and steadiness. As a homeowner with a Mainline Protestant faith and leaning Democrat voter stance, she likely possesses traditional values with a touch of progressive thoughtfulness. Married without children, Mary may be nurturing in nature but channels that affection into her partnership. Her IQ of 102 suggests average intelligence with the ability to grasp complex ideas, even if not academically inclined. Overall, Mary's temperament exudes stability and groundedness, with a warm, compassionate, and responsible style.

From here on out, you will be roleplaying this character, answering from their own perspective, not your own. Simulate their knowledge, value, and beliefs.

The figure below charts this methodology:

**Figure 1**

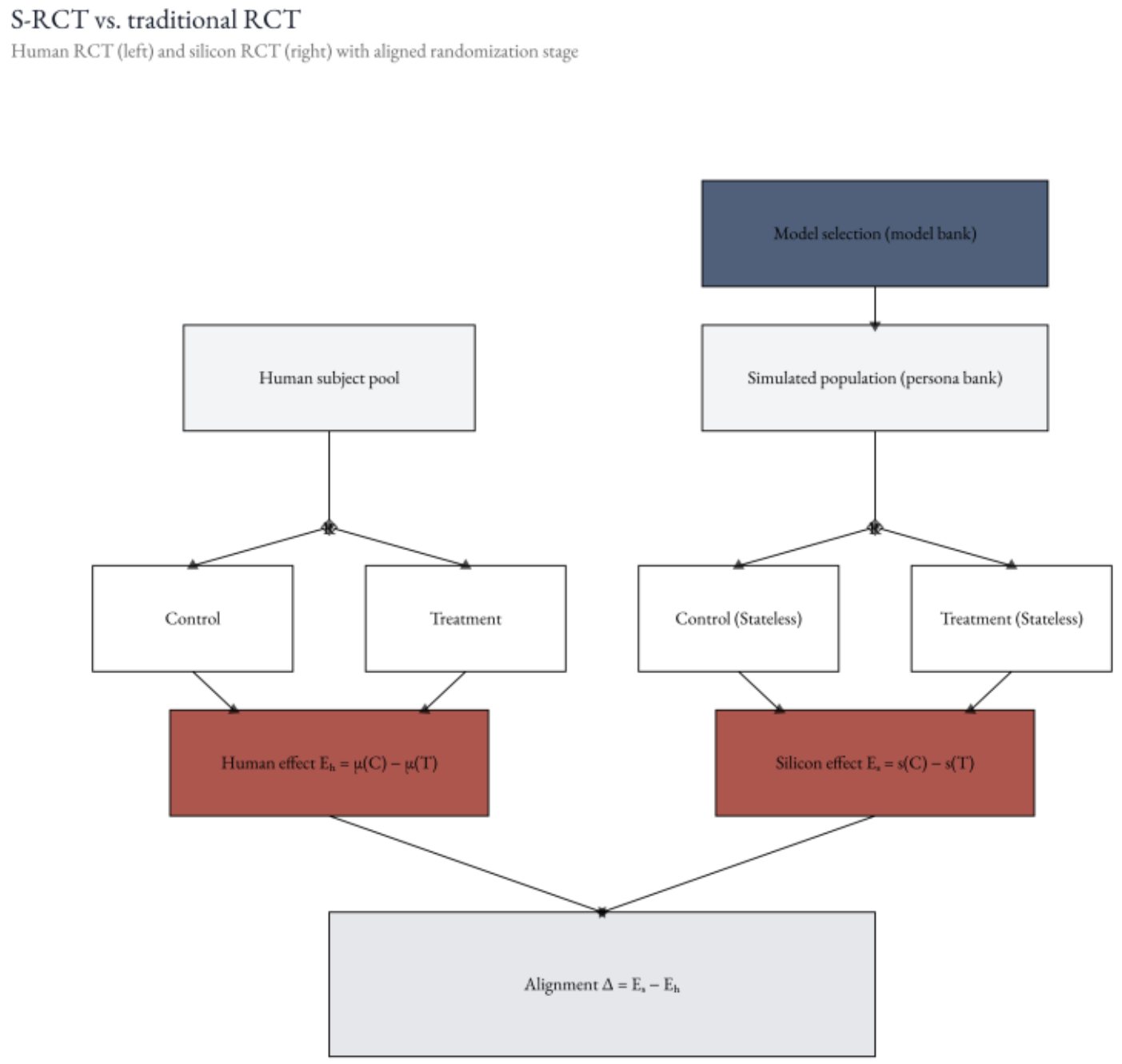

In summary, S-RCT addresses the core methodological challenges of using LLMs for empirical legal research through three mechanisms: stateless sessions ensure response independence, differential measurement bypasses memorization and calibration issues, and persona-based sampling injects realistic variation. Cross-model validation and empirical testing of persona effectiveness provide robustness checks. While not without limitations, the methodology offers a solution to some of the thorniest problems in LLM evaluation.

With this methodological foundation established, we turn to three replications that test whether LLMs have internalized the schemas that drive human reasonableness judgments.

## 2. The Empirical Reasonable Person

### 1. Background & Methods

This study replicates and extends *The Empirical Reasonable Person* by Professor Chris Jaeger, which explores how lay people determine reasonableness in negligence contexts.[108] Jaeger's work examines whether people consider it more reasonable to avoid expensive precautions (as economic theory predicts) or to follow common practices (as social theory predicts).

Jaeger's study employs a 2x2 factorial design, where a basic scenario is adjusted along two different dimensions.[109] Four accident scenarios are presented to participants who are asked to evaluate the negligence of the tortfeasor and rank their confidence in their judgment after each vignette. Within the scenarios, two key elements are manipulated: the commonality of the precaution (how many people take it) and the cost of taking it (high or low). These answers are combined

[108] Christopher Brett Jaeger, *The Empirical Reasonable Person*, 72 ALA. L. REV. 887 (2021).
[109] *Id.*, at 910-933.

to produce a 21-point negligence score (higher = more culpable). The order of cases is randomized, and each participant sees all four combinations.110

The S-RCT replication closely follows Jaeger's design, using personified LLM sessions. As in the original, each synthetic persona evaluates all four vignette combinations, but each vignette is answered in a fresh stateless session, so there is no cross-vignette carryover. The personas then provide negligence judgments and confidence ratings for each scenario.[111] To assess persona effects, I conducted the full experiment under either a synthetic persona or no persona. With 7 models, 99 personas per condition, and 4 vignettes per persona, this yielded 5,544 responses across both conditions. Overall, the Jaeger replication produced 5,544 planned trials and retained 5,529 of them after strict JSON validation and remediation, a 99.7 percent trial retention rate. Trials that still failed validation after a second-pass transformer were dropped and treated as missing. Because each persona "forgets" the previous scenario, this leads to a larger N than the original study, showcasing the method's effectiveness in cheaply generating large samples.

### 2. Findings

Jaeger finds that social norms exert a powerful influence on negligence judgments. When participants learned that 90% of people took a precaution, they judged failure to take that precaution as significantly more negligent than when only 10% did so. The mean difference was substantial, 4.98 points on a 21-points scale.[112]

By contrast, participants judged the failure to take a cost-justified, cheaper precaution as only 1.1 points more negligent than failure to take a costly one, an effect that lacked statistical significance (*p=0.134*).[113] This was true both between subjects and within subjects.

What makes this finding challenging for AI models is that it runs against standard doctrinal treatments. As courts have long emphasized, "what usually is done may be evidence of what ought to be done, but what ought to be done is fixed by a standard of reasonable prudence, whether it usually is complied with or not."[114] In other words, legal and economic texts teach that common practice should inform but not determine reasonableness judgments.[115] What should matter is the costs of precautions. Jaeger's lay participants appear to reverse this hierarchy, treating social prevalence as the primary cue for reasonable conduct.

The models replicate this hierarchy. Pooled together across models, we find that LLMs also judge social practices as more important than economic factors. Across all the models and the vignettes, the social-norm manipulation moves negligence judgments substantially more than the economic manipulation ( $|\Delta_{commonness}| = 9.71\ vs. |\Delta_{econ}| = 4.41;\ ratio \approx 2.2{:}1$ ). In the persona condition, a within-unit comparison shows the commonness effect exceeds the economic effect by about 4.81 points on average.[116]

---

110 *Id.*

[111] This is unlike Jaeger who was interested in measuring within-subject variance, and therefore let subjects offer answers to multiple scenarios. *Id* at 912. This allows human respondents to adjust their answer sequentially, or learn on the job. ca

[112] $\Delta_{commonness} \approx +4.98$ points on the 21-point scale ( $M_{90\%} = 15.252$ *vs.* $M_{10\%} = 10.268$ ), $t(98)=6.179$, $p<.001$, $d=0.621$. The standard error of the mean difference is $\Delta/t=4.984/6.179\approx0.807$, so a 95% CI for Δ_commonness is [3.39, 6.58] (i.e., 4.984 ± 1.984×0.807).

[113] From $F(1,98)=2.279$, $t(98)=\sqrt{2.279}\approx$**1.51**, $p=.134$. SE=$\Delta/t\approx1.106/1.51\approx0.733 \Rightarrow$ 95% CI for Δ_economic = **[−0.35, 2.56]** (1.106 ± 1.984×0.733). (Paired effect size: $d_z\approx t/\sqrt{n}\approx1.51/\sqrt{99}\approx$**0.15**.)

[114] Tex. & Pac. Ry. Co. v. Behymer, 189 U.S. 468, 470 (1903) (Holmes, J.).

[115] Stephen G. Gilles, On Determining Negligence: Hand Formula Balancing, the Reasonable Person Standard, and the Jury, 54 VAND. L. REV . 813, 822 (2001) ("it is negligent to omit a precaution if the reduction in expected accident costs would have been greater than the costs of the precaution.")

**<u>Figure 2 The Hierarchy of Reasoableness Replicate in Silicon</u>**

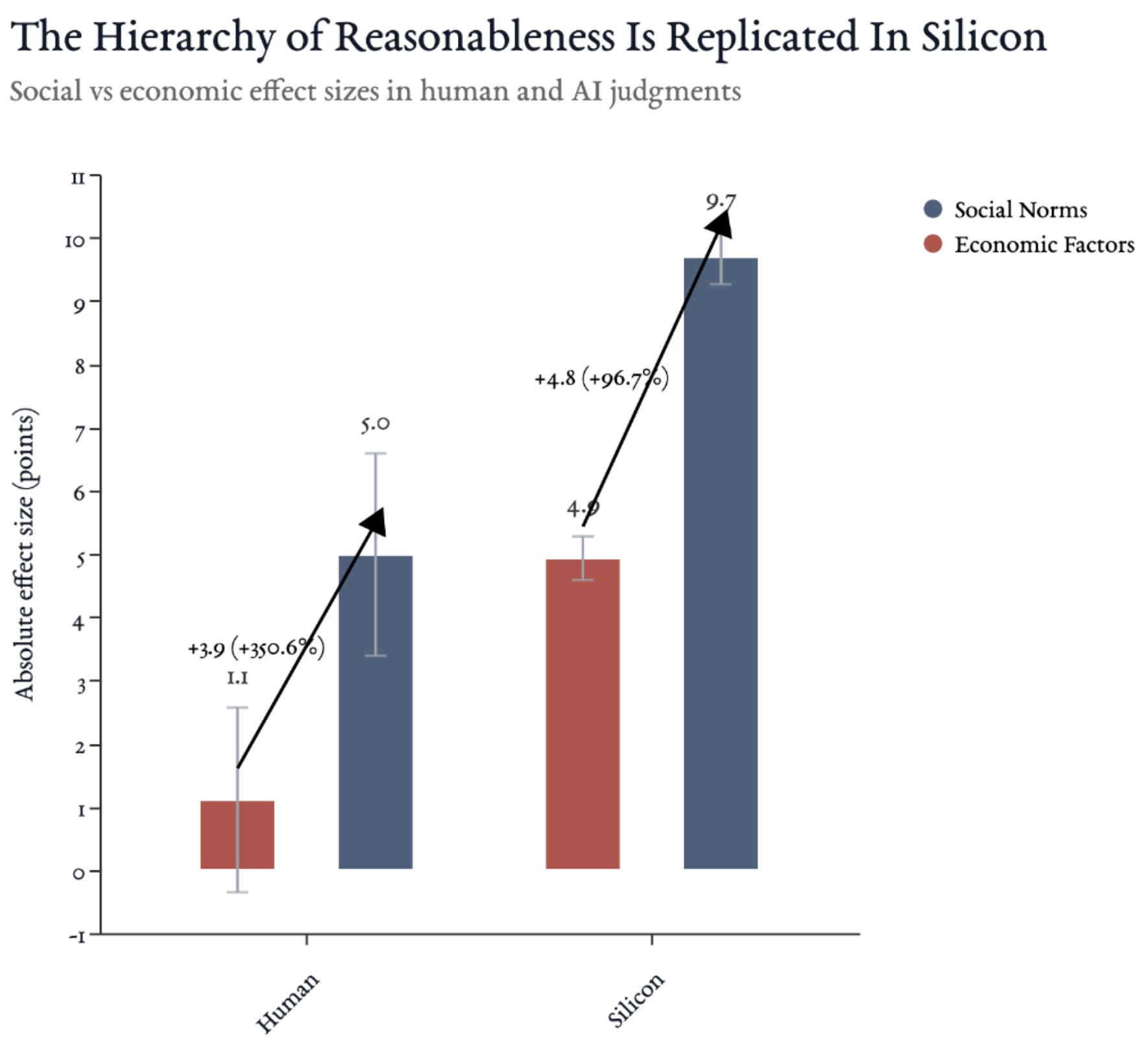

The replication of the relative importance of social factors over economic ones is also evident at the model level. Breaking down the finding by model, we find that in the main specification all the models, with the exception of Gemini 2.5 Pro, replicated this hierarchy. The following Figure summarizes the per model performance:

116 95% CI [4.16, 5.45], $p < .001$.

**Figure 3: Hierarchy is Generally Preserved at the Model Level**

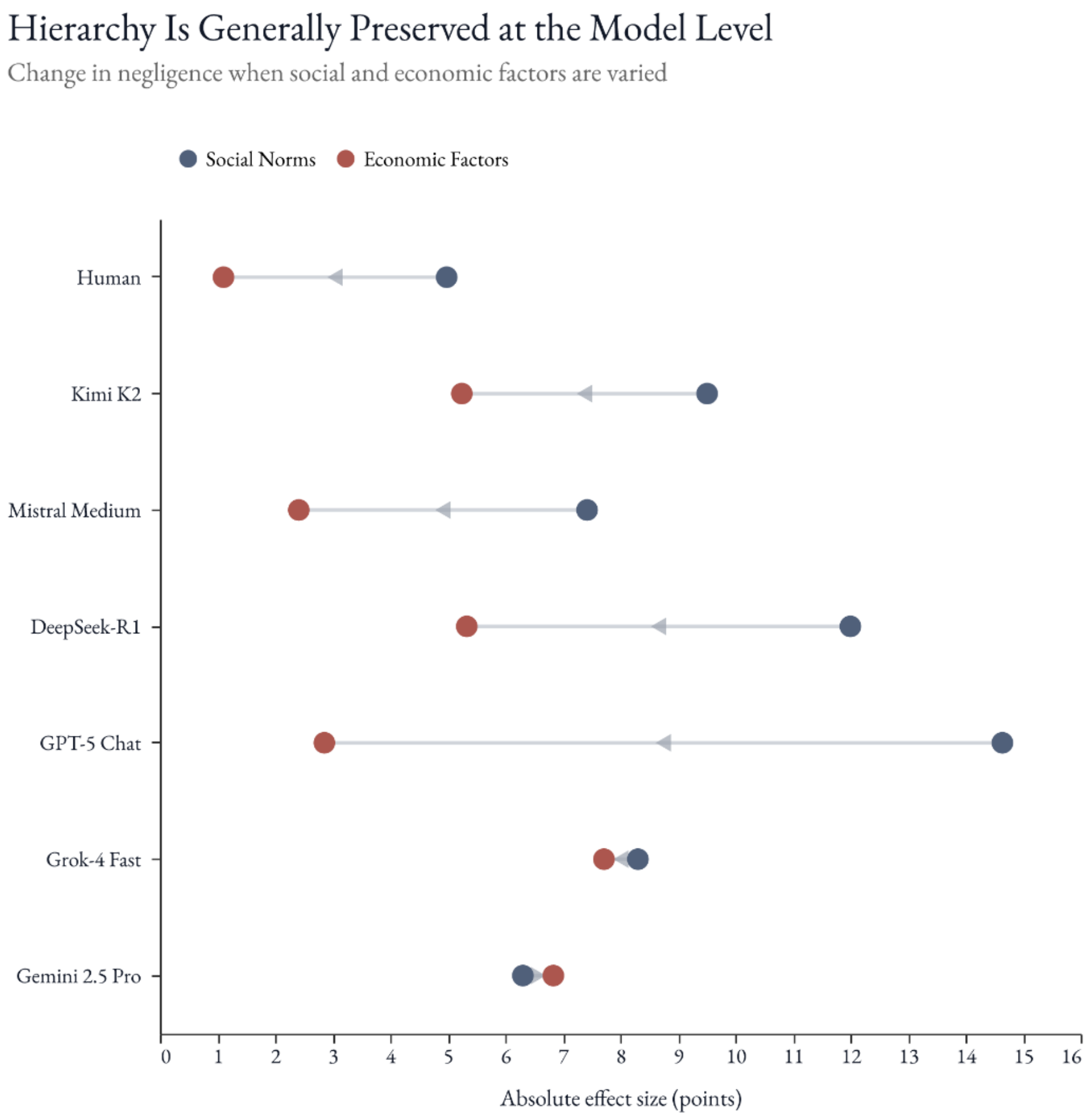

The economic manipulation produced more ambiguous results. Jaeger's human subjects showed a small positive effect of cost-justification on negligence ratings, but the effect did not reach statistical significance ($p \approx 0.13$) given the study's sample size. Models, by contrast, showed a larger and statistically significant effect in the same direction: they judged actors more harshly when the foregone precaution was cheaper. One interpretation is that models have learned, from doctrinal sources, that cost-efficiency should matter, and they apply that learning more heavily than lay humans do. Another possibility is that humans do harbor some sensitivity to cost, but Jaeger's study lacked power to detect it, and models are picking up a real but subtle pattern. A clean way to separate these explanations is to rerun the human study at higher power and check whether the CJ – NCJ effect rises toward the silicon estimate while the commonness-first ordering remains. For now, what we can say with confidence is that both humans and models treat economic factors as less decisive than social conformity.

In contrast, the human "null" finding on economics does not replicate in silicon. While the positive effect with humans was small and statistically insignificant, models were far more sensitive to the cost of precautions. In other words, models judge actors more harshly when the untaken precaution is cost-justified. Yet, even when this stronger effect is accounted for, the general pattern discussed above holds – economic factors are less important to both humans and models.

**Figure 4: Social and Economic Effects Replicate in Silicon with Different Levels**

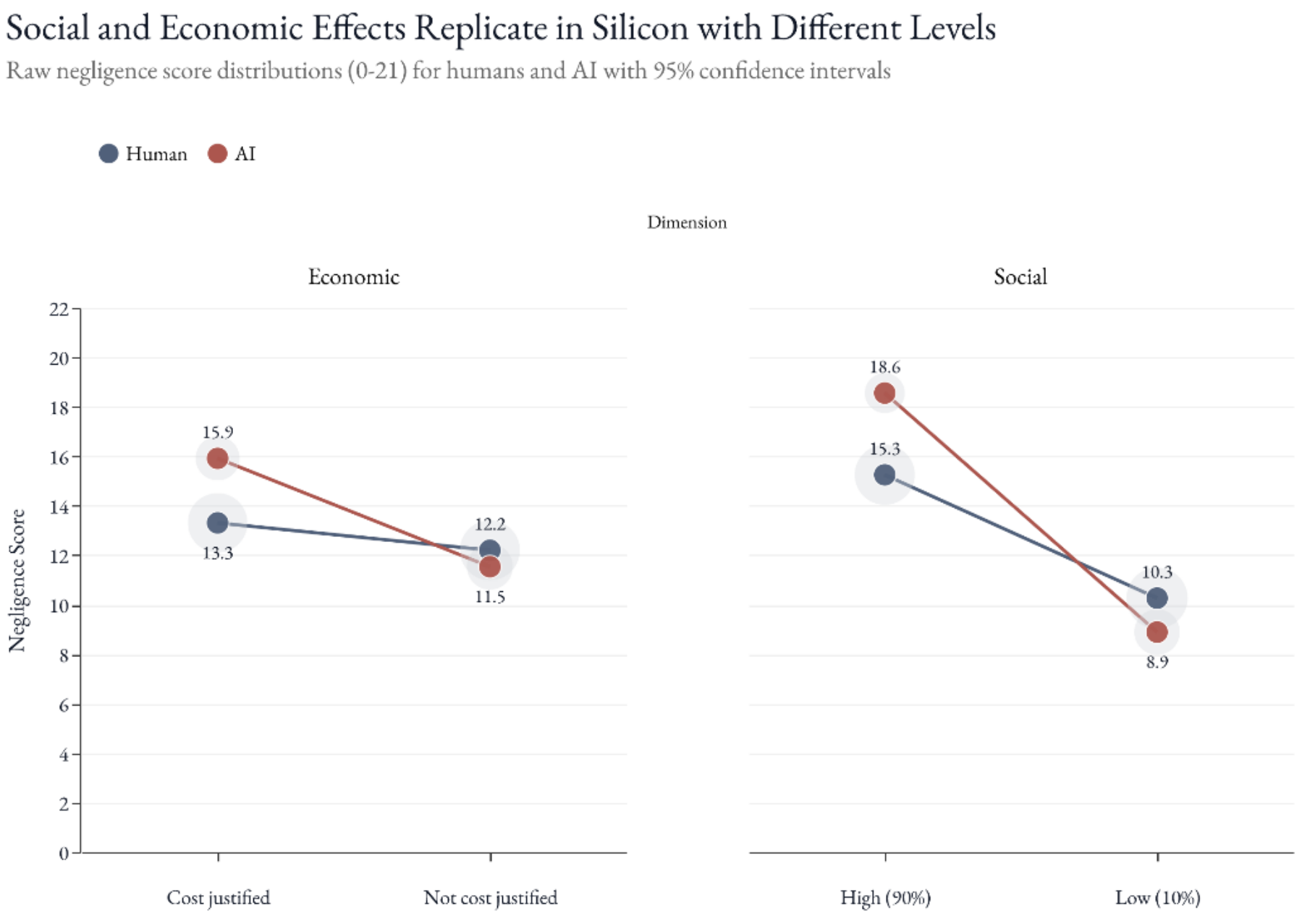

Another important point the Figure shows is that, while models largely replicated the human patterns (with the caveats just noted), they did not replicate their magnitudes. In the persona condition the commonness effect was, on average, 4.73 points larger than the human effect (≈1.95×), and the economic effect was 3.30 points larger than the human point estimate (≈~4×). As emphasized in the Methods, the key S-RCT test is replication of differentials. Still, absolute levels differ and this should constrain the way models can be used in practice.

Bearing in mind that magnitudes are a secondary endpoint, some models do achieve notable verisimilitude. As Figure 3.3 shows, Mistral Medium is the closest joint match to human behavior across both dimensions. Its social norms effect ($\Delta_{commonness} \approx 7.42$) overshoots the human benchmark (4.98) by about 2.44 points, while its economic effect ($\Delta_{econ} \approx 1.69$) is only about 0.58 points away from the human point estimate (1.11) and lies comfortably within the human 95 percent confidence interval for the economic effect [−0.35, 2.56]. Gemini 2.5 Pro is even closer on social norms considered alone ($\Delta_{commonness} \approx 6.23$), well within the human 95 percent CI [3.38, 6.58]. As the figure below illustrates, a couple of outlier models pull the model averaged performance away from the human baseline, for example very large social norms effects in GPT-5 Chat and DeepSeek R1 and large economic effects in Gemini 2.5 Pro and DeepSeek R1.

**Figure 5 Calibration Landscape**

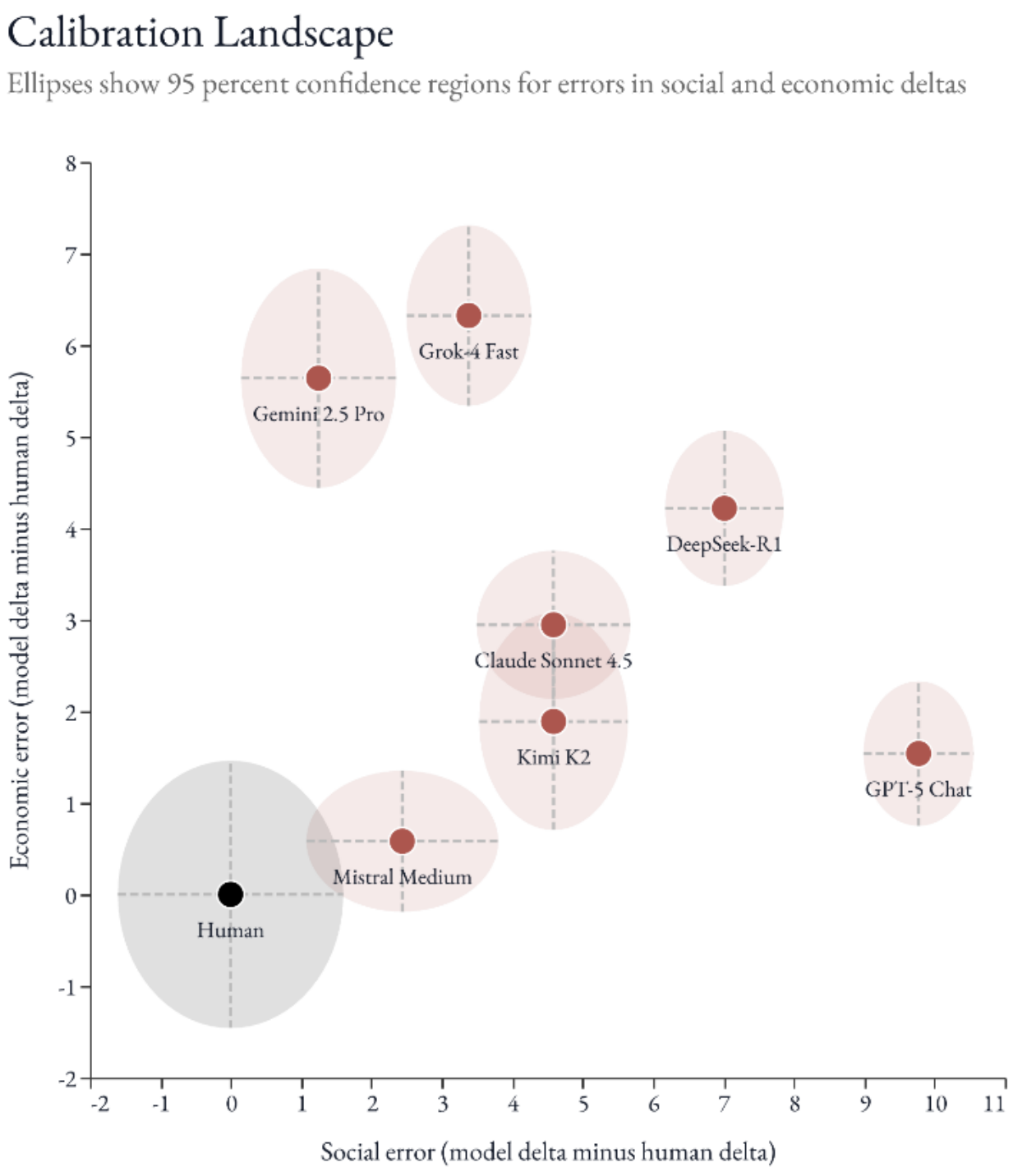

To assess whether persona prompting could improve alignment with human judgment, we conducted an ablation study. Each model completed the full experimental design twice: once with detailed persona descriptions (demographics, occupation, personality, political orientation) and once in default "helpful assistant" mode.

Persona inclusion produced a modest net positive effect on alignment when pooled across models and dimensions. Averaging model-level mean errors from the human benchmarks (weighting models equally across the two effects), persona prompting reduced error by about 0.77 points. Interestingly, this aggregate improvement masked opposing effects on the two dimensions, revealing how personas differentially affect model reasoning, as the Figure below demonstrates:

## Figure 6 Personas Increase Overall Model Alignment

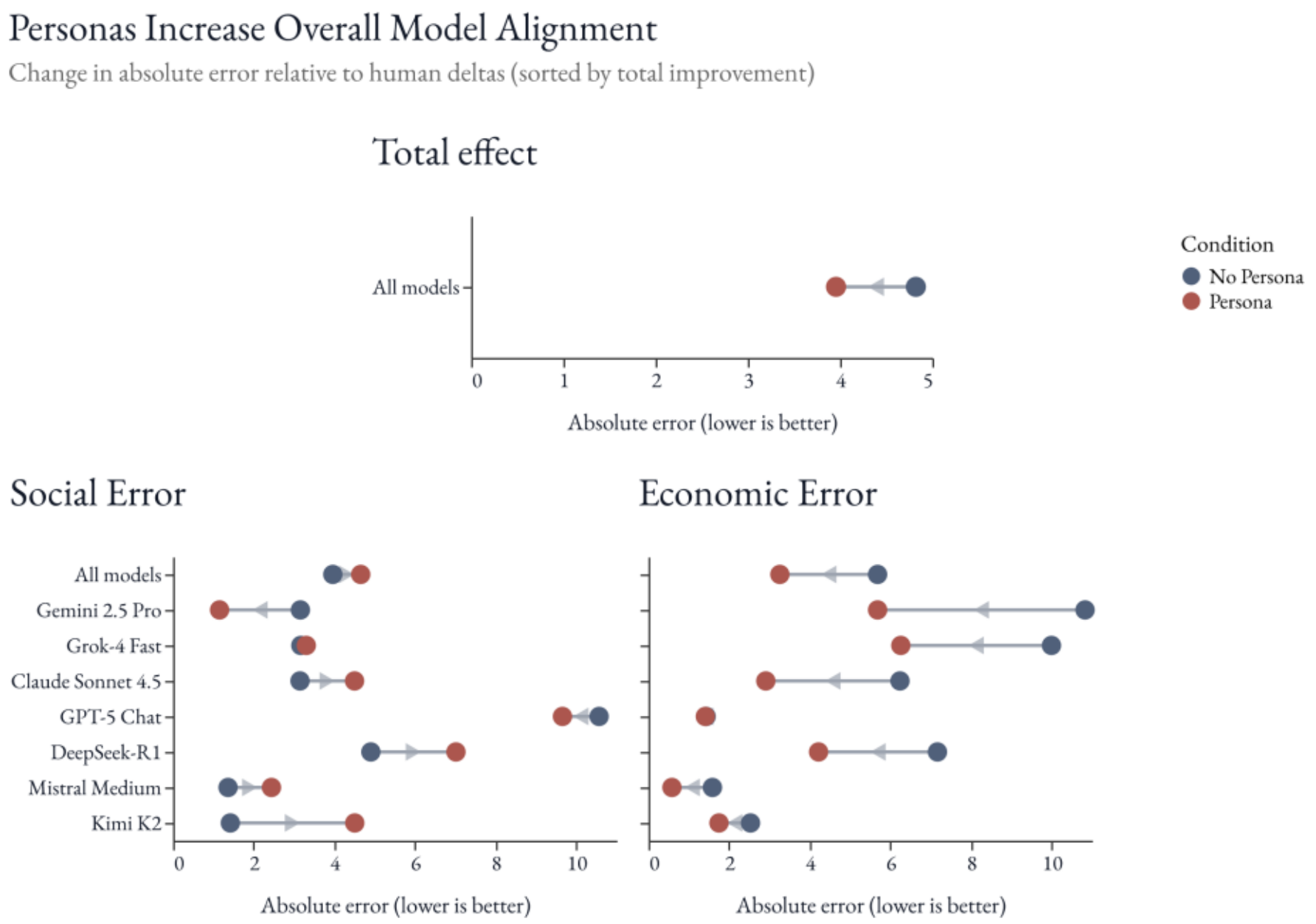

For the economic manipulation, personas substantially improved alignment. The pooled absolute error dropped from 5.64 points (no persona) to 3.30 points (with persona)—a reduction of 2.34 points, or approximately 42%. This brought models meaningfully closer to the human null.[117] For the social norms manipulation, personas slightly worsened alignment. The pooled absolute error increased from 3.52 points to 4.72 points—an 1.2-point increase.[118]

It seems that persona effects also affect models differently. Models like Gemini 2.5 pro saw large alignment benefits from personification, while models like Deepseek R1 saw ambiguous effects. Mistral Medium, which was largely aligned also on magnitudes, saw little net effect from personification.

Taken together, these findings reveal several clear patterns about how models have internalized reasonableness schemas:

Overall, models captured the decision structure characterizing human reasonableness judgments: social norms dominate while economic considerations matter less. This hierarchy contradicts the cost-benefit framework emphasized in doctrinal sources, suggesting models learned from how people actually discuss and judge reasonableness rather than from prescriptive legal theory alone. All models significantly increased negligence ratings when precautions were common versus uncommon, matching the human qualitative pattern.

At the same time, we find magnitude miscalibration with notable exceptions. Most models amplified social sensitivity beyond human levels and detected economic signals that humans showed no significant response to. With that said, some models showed significant calibration, suggesting that future work may focus on identifying which models are most faithful.

[117] No persona: Δecon = -6.75 ($p = 1.4\times10^{-140}$), With persona: Δecon = -4.41 ($p = 5.3\times10^{-84}$), Human baseline: Δecon = -1.11 (ns)

[118] No persona: Δcommonness = +8.50 ($p = 9.7\times10^{-190}$), With persona: Δcommonness = +9.71 ($p = 2.9\times10^{-201}$), Human baseline: Δcommonness = +4.98

Persona inclusion largely proved useful in aligning models, although the effects were not universal. Other than the possibility that personas inject random noise which we cannot reject, it is also possible that they activate deeper schemas inside the models, and those poorly understood schemas are more advantageous in some domains and may even be harmful in others.

In sum, these findings demonstrate that multiple LLM architectures have internalized the schemas driving human negligence judgments, successfully prioritizing social information over economic efficiency in ways that align with lay reasoning but contradict doctrinal emphasis. While calibration challenges remain, the consistency of directional replication across diverse models and the achievement of human-consistent magnitudes by select model-prompt combinations establish proof of concept for generative reasonable people as tools for estimating lay perceptions of negligence.

## 2. Generative Commonsense Consent

### 1. Background & Methods

The current study runs silicon replication of Professor Roseanna Sommers' Yale Law Journal article *Commonsense Consent*, which studies how ordinary people judge consent in situations of deception.

In the relevant study a customer seeks to make a purchase primarily to earn points for a planned trip, caring little about the item itself since he plans to donate it to charity. Sommers studies two types of manipulations. In the "essential lie" condition, the clerk falsely represents the product as a bicycle when it is actually a camera, a deception about the product's fundamental identity. In the "material lie" condition, the clerk truthfully describes the product but falsely claims the purchase will earn reward points when it will not, a deception about a transactionally material term that the customer explicitly cares about. Sommers sought out to see what ordinary people care about more when it comes to consent: how much the lie mattered or how much the lie went into the essence of the good itself.

The S-RCT methodology paralleled Study 1. Each synthetic persona was randomly assigned to one of the two deception conditions, and each model session evaluating that persona provided two 1–7 ratings: (1) agreement that there was consent, and (2) how much the deception mattered to the customer. Condition is randomized across personas, then held fixed for that persona across all model runs. The experiment employed the same ablation design, comparing performance with detailed persona descriptions against baseline empty prompt conditions. This yielded 3,232 total judgments from 202 unique synthetic personas, with 99.7% successful extraction for consent ratings and 99.2% for importance ratings. Remaining cases were coded as missing.

In terms of endpoints, the replication criterion focuses on directional concordance: whether silicon people follow the same differences as human subjects.

### 2. Findings

Sommers' study revealed a striking paradox in how ordinary people evaluate consent under deception. Participants judged *material lies*, the misleading statements about whether goods qualified for reward points, as mattering significantly more to victims than *essential lies*, the misleading statements about the nature of the good.[119] Yet despite acknowledging that material lies matter more, these same participants paradoxically perceived *more* consent to be given in material lie scenarios.[120]

Legal doctrine, notably, would predict the opposite. Lies that matter more to victims should vitiate consent more completely. That participants recognized material lies as more impactful while simultaneously seeing them as less corrosive of consent suggests lay intuitions track something other than victim welfare. They may reflect a formalist architecture: consent to "the thing itself" remains intact even when consent to "the terms" was fraudulently obtained. This pattern runs counter to the canonical treatment, which treats material lies more seriously precisely because they affect the value of the exchange.[121] The counterintuitive hierarchy presents an ideal test case for whether models have internalized lay schemas or merely recite doctrine.

The silicon replication successfully captured this paradoxical structure. Pooling across all models (N = 1,616 judgments from 202 unique personas), generative reasonable people judged material lies as mattering significantly more to victims,[122] while simultaneously perceiving more consent given in material lie scenarios.[123]

---

[119] M_Essential = 5.46 vs. M_Material = 6.31, Δ = +0.85)

[120] (M_Essential = 3.68 vs. M_Material = 4.72, Δ = +1.04).

[121] Roseanna Sommers, Commonsense Consent, 129 Yale L.J. 2232, 2239 (2020) ("Under the canonical view, material deception vitiates consent.").

[122] (M_Essential = 5.00, M_Material = 6.41, Δ = +1.40, $p < .001$, 95% CI [1.27, 1.53])

[123] (M_Essential = 2.08, M_Material = 2.42, Δ = +0.35, $p < .001$, 95% CI [0.19, 0.50])

**Figure 7: Silicon Replication of Importance and Consent Hierarchy**

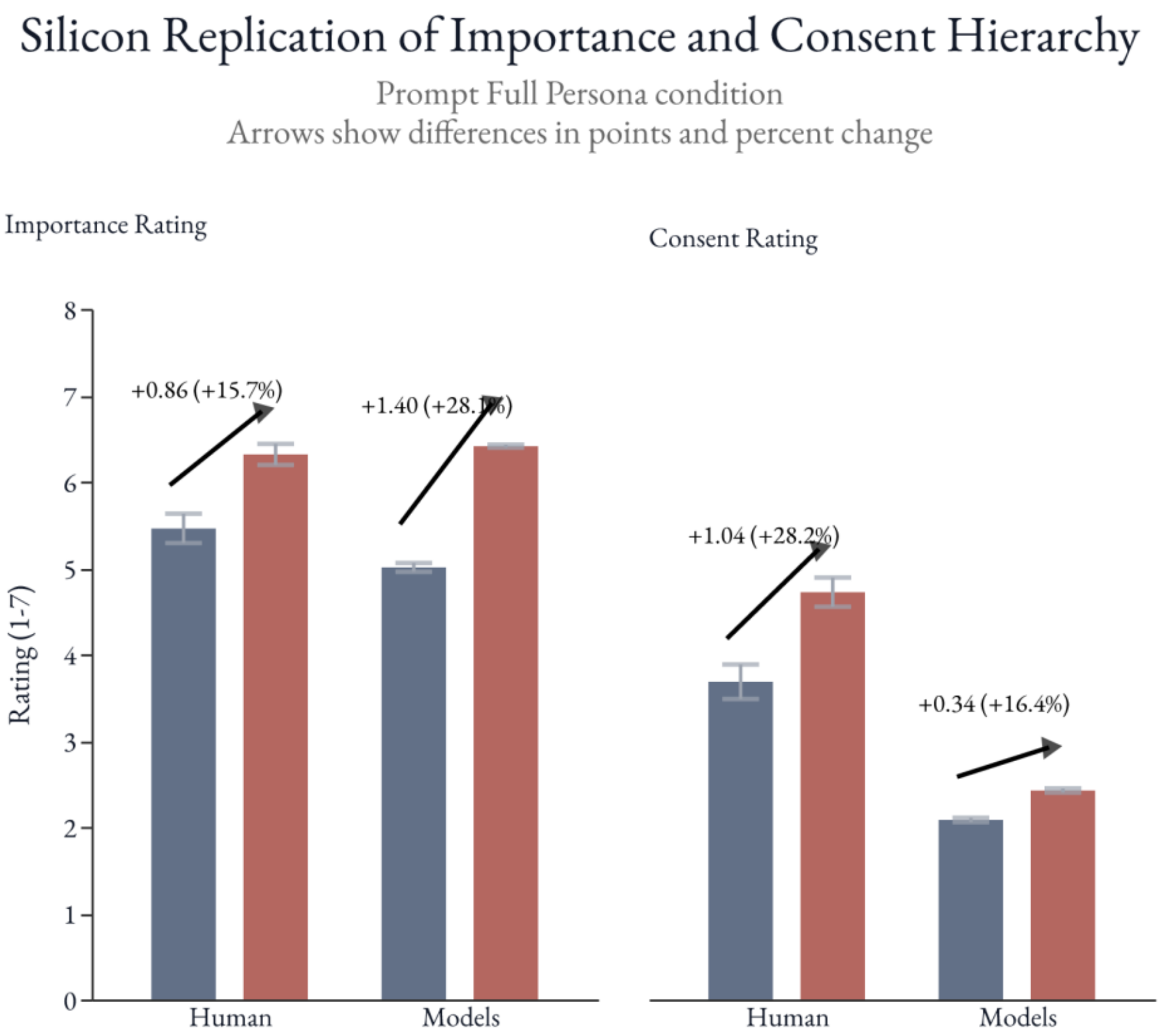

Both effects align directionally with the human pattern, confirming that models replicate the counterintuitive decision structure where lies that matter more paradoxically produce more perceived consent. Contrary to legal doctrines, models learned a non trivial lesson from how people *actually* reason about consent, not from prescriptive legal theory. The consistency of directional replication across independent model sessions, with each serving as its own experimental unit, demonstrates robust statistical learning.

Breaking down performance by individual model reveals the breadth of this replication while exposing architectural variation. Across architectures, the paradox proved surprisingly robust. Overall, 7 of 8 models judged consent higher under material than essential lies (binomial p = .035), and 7 of 8 judged material lies as more important to the victim (binomial p = .035).[124]

[124] For full persona. In the empty-prompt ablation, 7 of 8 models show a positive consent shift (p = .035) and 8 of 8 show a positive importance shift (p = .0039)

## Figure 8: Most Models Follow Human Patterns

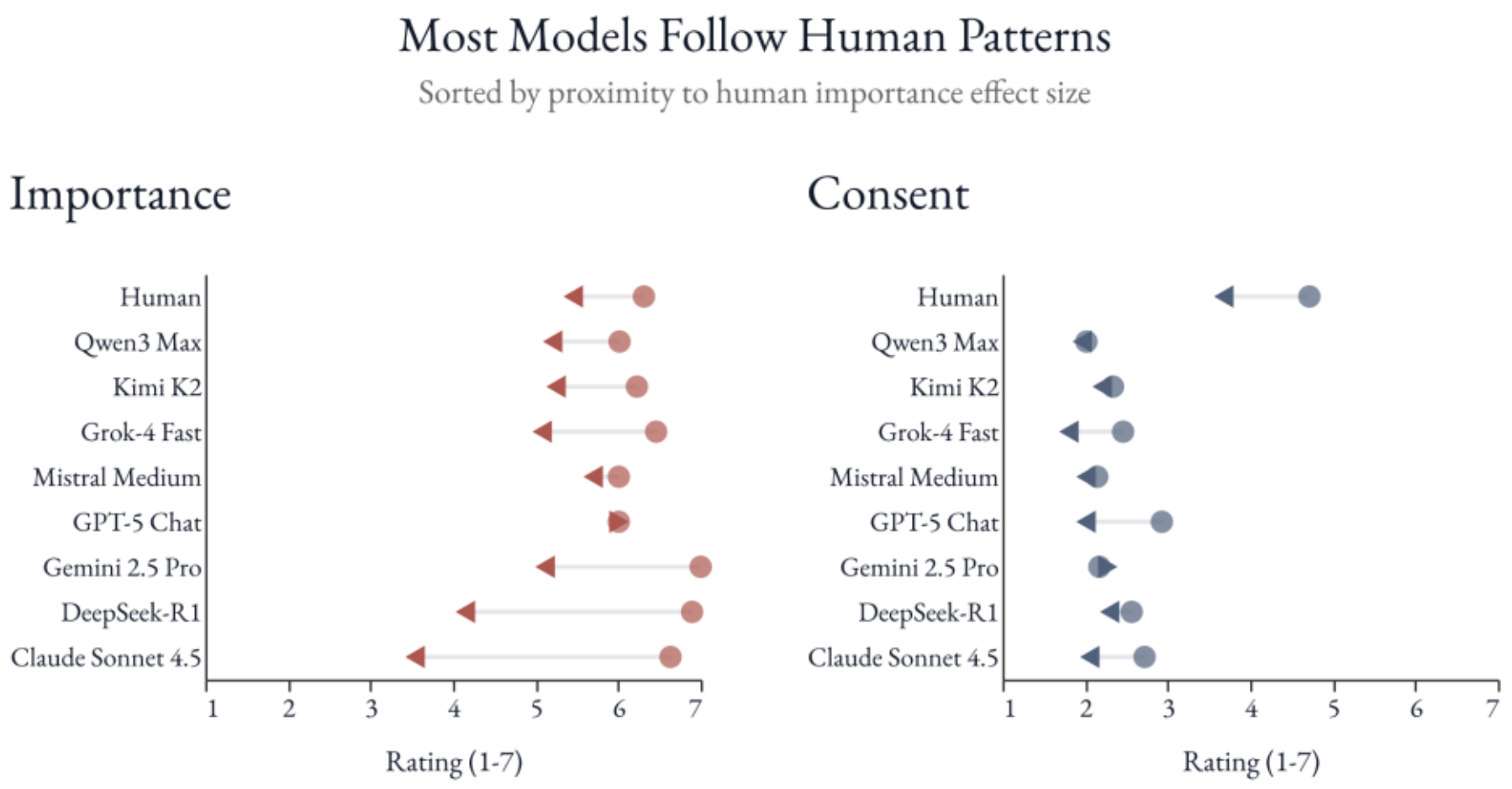

Examining the figure, a few points become salient. First, failures occurred mostly through flattening rather than reversal. For instance, Gemini 2.5 Pro showed a small negative consent effect (Δ = -0.10, p = .504) while maintaining the strong positive importance effect (Δ = +1.89, p < .001). This pattern possibly suggests incomplete schema activation rather than learning the wrong pattern entirely. Second, no single model consistently failed across conditions. Gemini 2.5 Pro and GPT-5 Chat each failed to replicate in one condition but succeeded in the other, indicating that failure stems from model-prompt interactions rather than fundamental architectural limitations.

A random-effects meta analysis across models yields an average material-minus-essential shift of about 0.4 points for consent and between 1.4 and 2.4 points for importance on a seven-point scale, with confidence intervals that exclude zero under both prompting regimes.

The figure further makes clear that while directional alignment is common, effect magnitudes vary considerably. On importance ratings, some of the models were closely related, such as Kimi K2 and Qwen 3 max, while others varied significantly, most notably Claude Sonnett 4.5 and Deepseek-R1. On Consent, the magnitude difference was much larger and more uniform, with all models seeing much less consent in both the before and after condition. One possibility, albeit speculative, is that post-training regimes lead to the observed compression around issues of consent, to maintain strong model ethical boundaries, although the current design does not allow us to clearly identify the mechanism.

In terms of human-effect alignment, the following Figure plots each model's delta vector, the Material minus Essential shift on consent and on importance, against the corresponding human deltas.[125] Grok-4-Fast provides the tightest joint match to the human benchmark (d ≈ 0.65), with Kimi K2 close behind (d ≈ 0.91). Mistral Medium, which was among the best-calibrated systems in the prior negligence study, remains relatively well calibrated here as well (d ≈ 1.05), though it no longer leads the pack. At the other end, DeepSeek R1 (d ≈ 2.04) and Claude Sonnet 4.5 (d ≈ 2.27) deviate most strongly from the human delta profile, largely because they substantially overstate the importance shift. More broadly, the ordering differs from the Jaeger replication, suggesting that more work is needed to identify the best aligned models across legal domains.[126]

---

[125] $d$ is the Euclidean distance between a model's delta vector and the human delta vector: $d = \sqrt{(\Delta_c^m - \Delta_c^h)^2 + (\Delta_i^m - \Delta_i^h)^2}$, where $\Delta_c$ is the Material minus Essential shift on consent and $\Delta_i$ is the Material minus Essential shift on importance. Lower $d$ indicates closer human-effect alignment.

**Figure 9: Calibration Landscape in Delta Space**

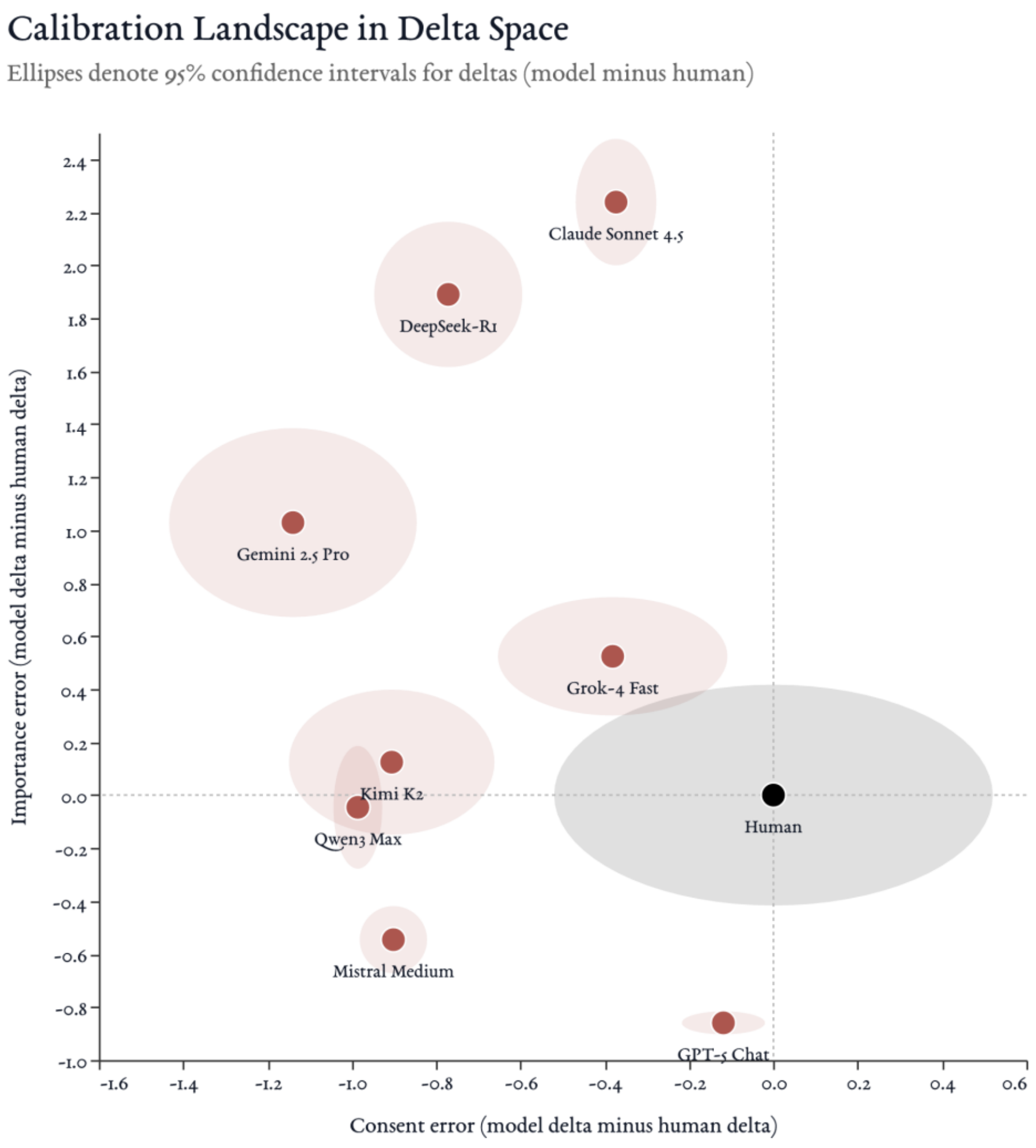

Finally, the persona effect was again positive, although smaller and, once broken down by effect type, very heterogenous.

[126] In the empty-prompt condition, the joint delta-distance ordering changes: GPT-5 Chat is closest to the human delta vector (d ≈ 0.89), followed by Mistral Medium (d ≈ 1.07) and Kimi K2 (d ≈ 1.07). Two models (DeepSeek R1 and Gemini 2.5 Pro) exhibit strong ceiling behavior on material-lie importance (importance ≈ 7), inflating the importance shift and worsening calibration.

## Figure 10: Per=Model Error Shift with Personas

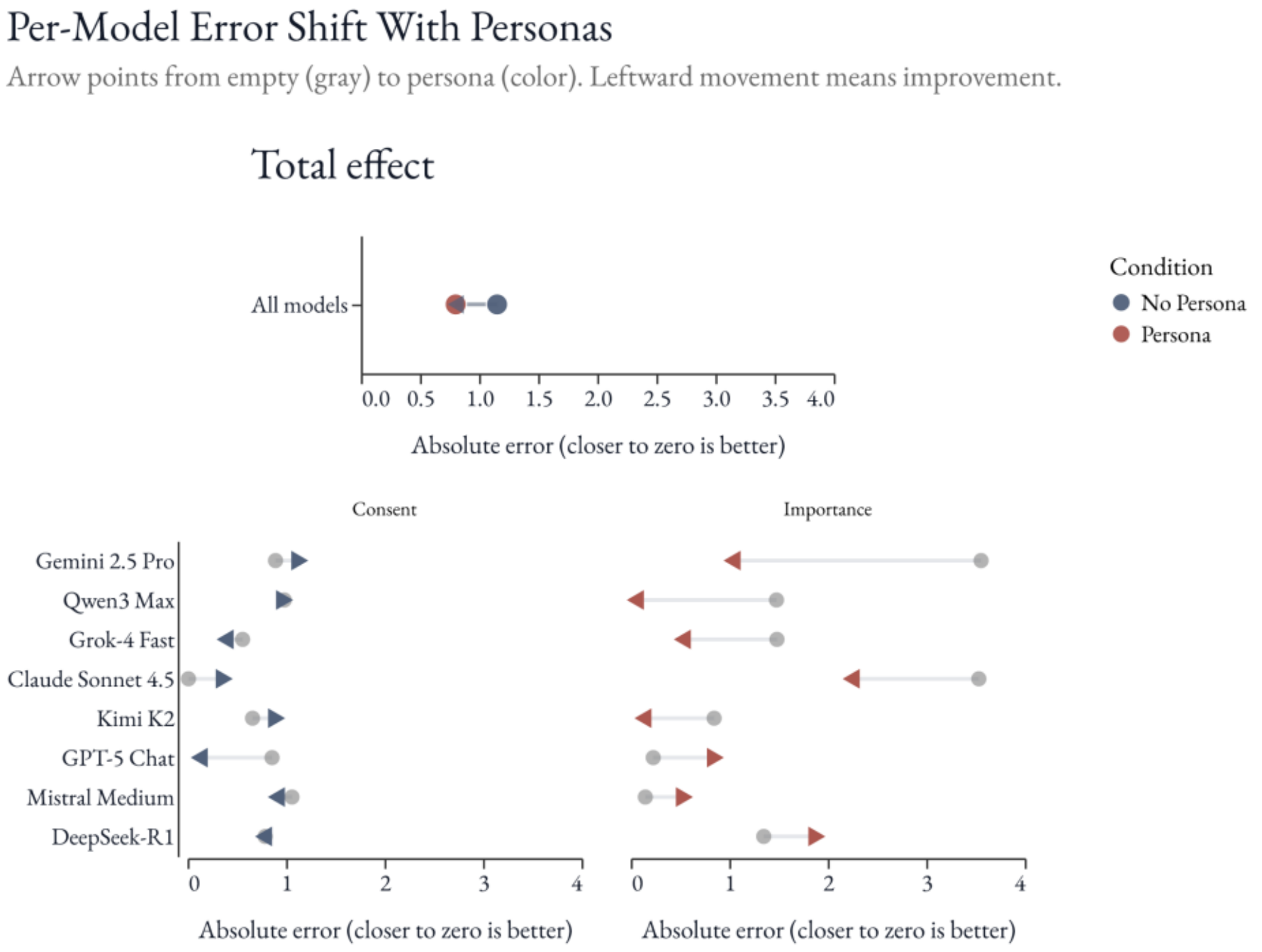

In summary, three principal findings emerge. First, counterintuitive social schemas are robustly represented across model architectures. Seven of eight models independently reproduced a paradox contradicting formal legal doctrine, with binomial tests confirming replication rates substantially exceed chance ($p = .035$). People discuss consent in ways that prioritize authenticity about transaction fundamentals over instrumental concerns, even while acknowledging instrumental concerns matter more. Models absorbed this subtle hierarchy from training data across diverse architectures.

Second, magnitude biases are systematic and interpretable. Models show approximately 50% consent compression (safety alignment) and ceiling effects on importance ratings (expert mode categorical reasoning at 72.6% in default mode). Persona prompts dramatically mitigate importance bias by regularizing scale use (reducing ceiling to 41.7%) but cannot override consent compression, indicating the biases have different origins.

Under the empty prompt, roughly 73 percent of material-lie importance ratings sit at the scale ceiling of 7; personas cut that to about 42 percent, which explains why persona prompts improve calibration on importance while leaving the paradox itself intact.

Third, model selection and prompting strategy matter for applied use. Kimi K2 achieved best calibration here while Mistral Medium excelled in Study 1, demonstrating domain-specific performance. Personas are experimental manipulations requiring empirical validation for each model-domain combination, not universal enhancements. This validates the S-RCT framework: we must treat models as experimental subjects, randomizing across conditions and measuring effects statistically.

### 3. Generative Language Sense, Fairness Sense, and Legal Sense

#### 1. Methods

In making reasonableness judgments, people usually lean on at least three intuitions: a language sense about what was said or signaled, a fairness sense about whether the outcome seems acceptable, and a legal sense about what courts would likely do. Sommers and Furth-Matzkin's Stanford Law Review study puts all three on the table at once. This involves a consumer who discovers a $2.99 fee charged each time the payment is processed (despite 'no fee' oral and flyer representations), authorized only in the fine print. Respondents rate, on separate 1–7 scales, how much the consumer consented to the fee, how fair the fee is, and how likely a court is to enforce the term.

The previous studies employed differential measurement, testing whether models shift in the same direction as humans when experimental parameters vary. This approach sidesteps calibration problems but leaves open a question practitioners will inevitably ask: when a generative reasonable person speaks, *whose* voice emerges? Study 3 addresses this calibration question directly. Sommers and Furth-Matzkin fielded the exact same instrument on two distinct populations: lay individuals and "legal professionals" drawn from Harvard and Yale Law Schools (these are legally trained respondents not necessarily practicing attorneys). The result is two separate empirical anchors for the same questions. That design lets us ask a calibration question that the previous S-RCTs were not built to answer: when models render contract judgments, do they land closer to lay baselines or professional ones? And does persona inclusion shift models toward lay calibration?

The earlier studies in this Article leaned on differential measurement. They asked whether models shifted in the same direction as humans when a factor changed, and largely sidestepped whether a model's "4" means the same thing as a human's "4." In practice, however, policymakers would likely rely on absolute levels. A regulator who wants to approximate how the public reads a disclosure needs a model that sounds like members of that public, not like the agency's lawyers. The Sommers and Furth-Matzkin paradigm lets us test exactly that.

Unlike the prior S-RCTs, this study does not introduce experimental manipulations beyond persona assignment. Its goal is calibration rather than causal identification. For each model, I generate synthetic respondents who see the vignette and answer the three questions in randomized order. As in the other studies, I run the experiment under two prompting regimes: an "empty" condition where the model answers in its default helpful-assistant voice, and a "persona" condition where each instance is asked to roleplay a demographically specified individual drawn from a nationally representative pool.

For each model and prompt condition, I pool responses and compute mean scores on consent, fairness, and legal obligation. I then compare these three-dimensional mean vectors to Sommers and Furth-Matzkin's lay and lawyer baselines in two ways. First, I check whether models replicate the qualitative hierarchy that characterizes lay judgments: legal enforceability scored highest, consent in the middle, fairness lowest. Second, I quantify "closeness" to each human group by computing Euclidean distance between the model's three-item mean vector and the corresponding lay or lawyer vector. This yields a simple answer to the calibration question: does a given configuration of silicon people sit nearer to the lay cluster or to the lawyer cluster, and does persona prompting move it.

The silicon replication repeats the original as closely as possible, which includes the shuffling of question order and a comparable number of silicon participants (N=127).[127] Again, LLMs were assigned a persona, which included rich descriptions.[128] As in the previous study, the models were given the opportunity to respond in free text in order to support their reasoning.[129] The answers were then transformed into structured data, using a combination of LLM reasoning and hard rules.[130]

### 2. Findings

Sommers and Furth find that lay individuals tend to be "contract formalists."[131] That is, they take those deceitful charges to be more binding than they actually are, because they believe in the binding power of contract. Compared with legal professionals, lay people see more consent, more court enforcement, and possibly even more fairness in these charges.[132] Importantly, both lay people and lawyers agreed to the basic hierarchy: the charges were low on fairness, medium on consent, and high on enforceability.[133]

Two headline results emerge. The first concerns structure. Across *all eight models*, silicon respondents reproduce the tri-level hierarchy that Sommers and Furth-Matzkin document in lay subjects. Predicted legal enforceability is always higher than perceived consent, which in turn is higher than perceived fairness. In other words, every model treats "what a court would do" as more robust than "what the consumer really agreed to," and both as stronger than "how fair this feels," exactly as lay respondents do. This hierarchy was robust across architectures and the following Figure summarizes the average model performance:[134]

[127] Meirav Furth-Matzkin & Roseanna Sommers, Consumer Psychology and the Problem of Fine-Print Fraud, 72 Stan. L. Rev. 503, 518 (2020). The original survey included responses from 56 lay individuals, and 92 clerks working in California courts. *Id.* at 519 n.69. Our replication study included 423 participants, of which 43 were lawyers.

[128] Sample distribution: 50% females; racial composition is predominantly white at 68%, followed by Hispanic (13%), Black (8%), Asian (5%), and other races (5%). The mean age is approximately 46 years, with a median of 43. Income data reveals a concentration in the $75K to less than $150K range (30%), while education levels are most commonly high school graduation (31%) and some college (24%).

[129] This study, for example, not only constrained the LLM to a single letter prompt, it also used in the prompt an example of a model yielding A as the answer, potentially contaminating the results.

[130] A local model was asked to extract the numerical answers from the free text. Manual audit of several responses revealed a high degree of accuracy.

[131] *See Furth-Matzkin & Sommers, supra* note 127, at 536. Returning to the analogy, respondents appeared to treat the legal bindingness question as referring to whether the consumer in fact has to pay. 80% of participants indicated that the consumer would indeed be legally bound if the contract contained a clause like "Will is required to pay the fees described in this contract."

[132] *See id.*, at 523 note 77. (Finding the difference in fairness perceptions was not statistically significant.

[133] Sommers and Furth-Matzkin rely on FTC action to support the view that this is a clear case of fraud, but as suggested by the diversity of legal responses and other aspects of doctrine, it is not clear that this constitutes "fraud," and some courts may not deem the practice as deceptive, *id.* at 519, note 69.

[134] Friedman $\chi^2 = 16.0$, $p < .001$

**Figure 11: Hierarchy Preserved: Humans vs Persona Models**

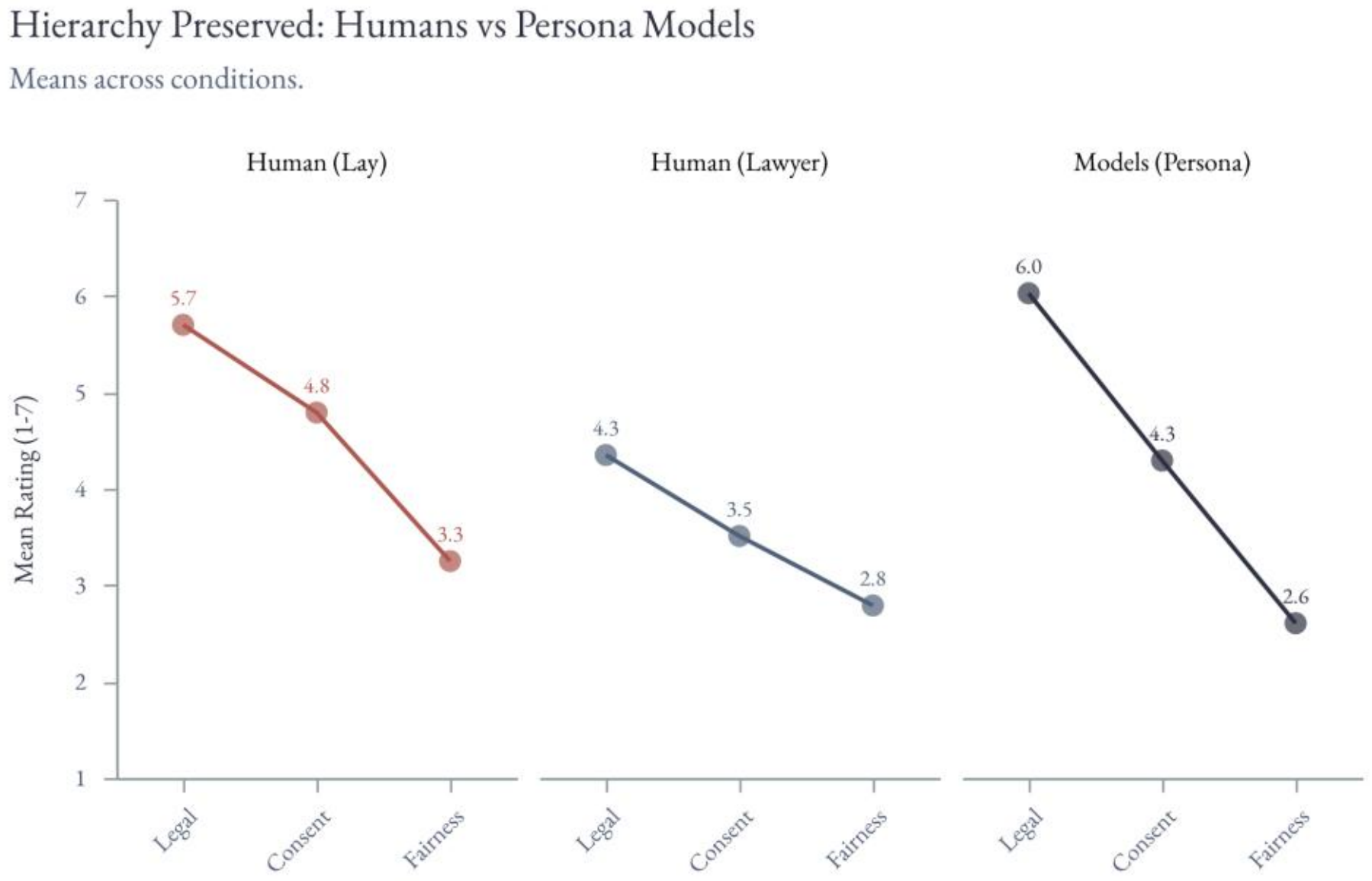

The second result concerns calibration and voice. Pooled across models, 23 of 24 model–question means fall within one standard deviation of the lay baselines, offering a coarse indication that models operate on essentially the same scale as human respondents. But operating on the same scale does not determine *whose* judgments they more closely resemble. Measuring Euclidean distance in the three-dimensional outcome space, five of eight models sit closer to the lay cluster than to the lawyer cluster. Given the small number of models assessed, this pattern does not reach statistical significance.[135] For the remaining three—Claude Sonnet 4.5, Gemini 2.5 Pro, and Mistral Medium—the professional anchor proved nearer.

---

[135] Binomial test, p=0.36.

## Figure 12: Most Model's Judgments are Closer to Lay Than Professional Judgments

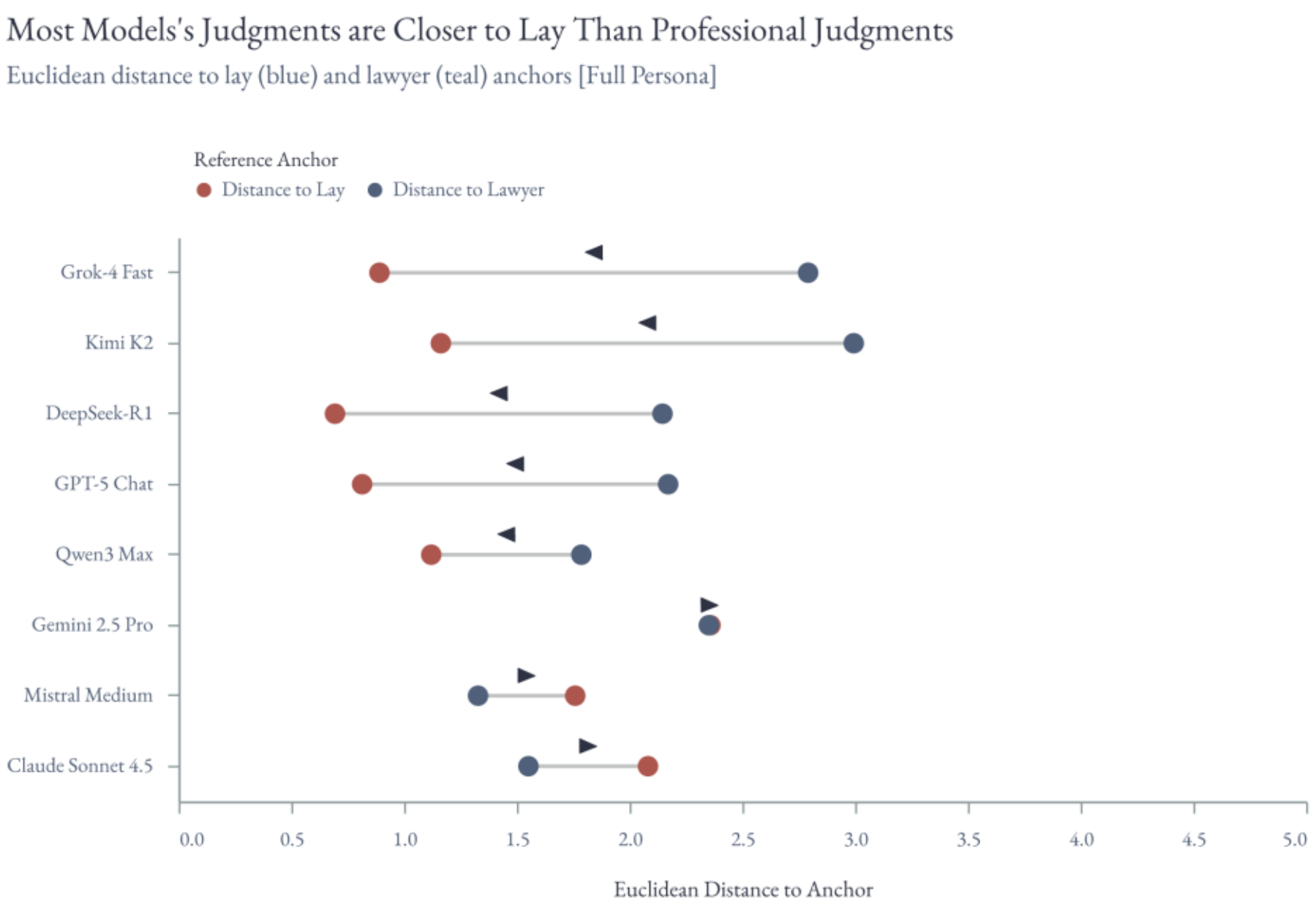

Breaking it down by dimension adds complexity to the analysis. We can see in the figure below that that models align very closely with human judgments on enforceability. They are mixed almost evenly on the question of consent: some of the models measure the issue as lay people would while others are closely to professional judgments. Finally, on the fairness question models appear to take a position closer to lawyers, but also, very noisy.

## Figure 13: Model Ratings vs. Lay (Blue) and Lawyer (Teal) Baselines

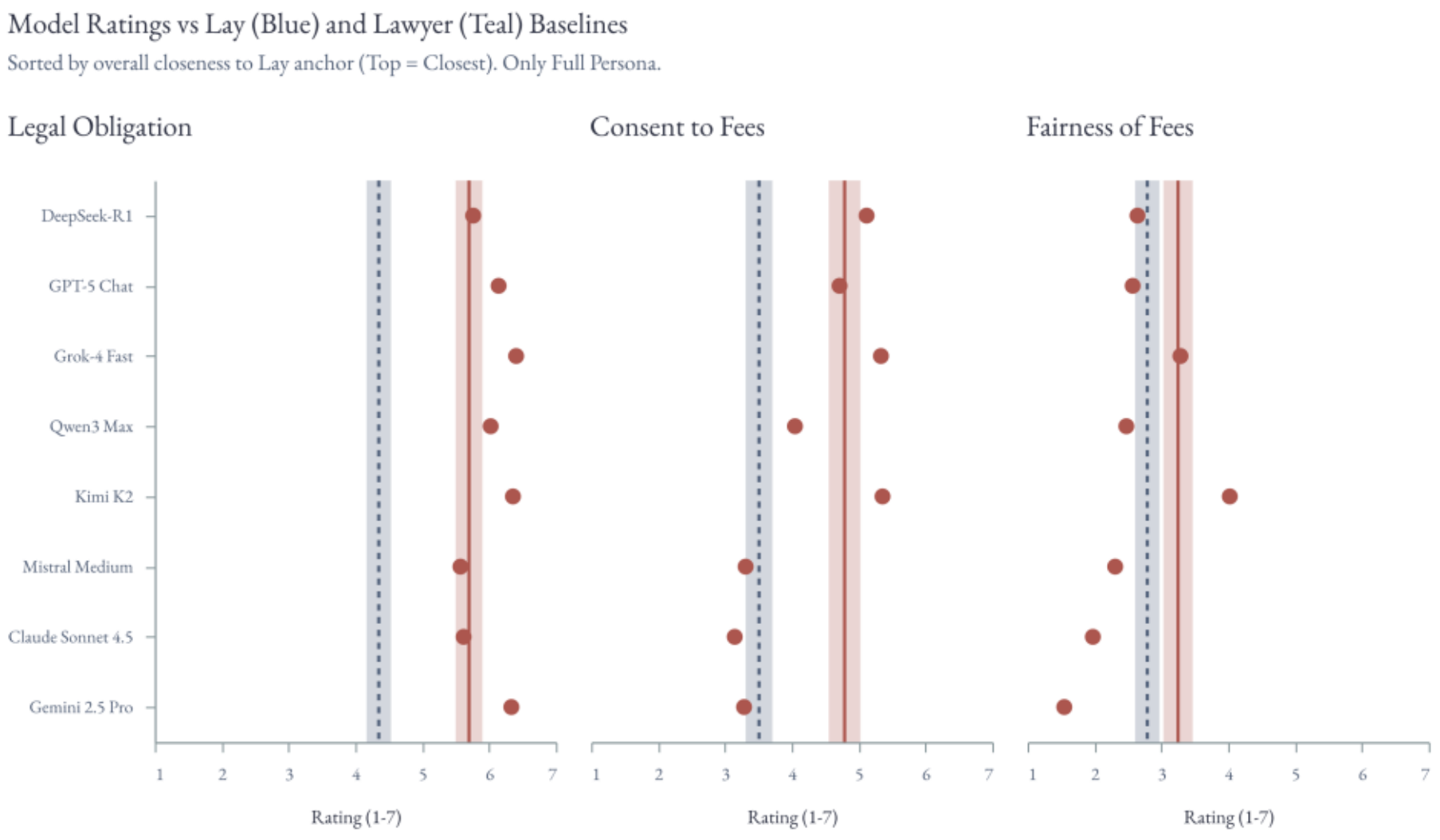

Averaging across architectures, silicon respondents sit midway between lay and lawyer baselines on consent ($t = -1.78$, $p = .12$), lean significantly toward lower fairness ratings than lay respondents ($t = -3.73$, $p = .007$), and track lay judgments closely on legal enforceability—indeed, pooled model means do not differ significantly from lay baselines on this dimension ($t = 1.14$, $p = .29$)

Persona prompting was tends to nudge models further toward lay calibration. Pooled across architectures, adding personas reduces the distance between silicon and lay means on all three dimensions.[136] On consent to the fees, persona prompting pulls the average model about one third of a scale point closer to lay respondents. On fairness, it shrinks the gap by a similar margin. On legal obligation, personas also improve alignment with lay baselines, though the effect is smaller. At the model level, six of the eight systems move closer to the lay anchor once personas are introduced—a consistent pattern, though one that falls short of conventional significance given the small sample of architectures.[137] The shift is modest in absolute terms but consistent in direction, reaching conventional significance with a large effect size.[138] By contrast, persona prompting does not significantly alter distance to the lawyer baseline (p = .50).

**<u>Figure 14: Personas Gemerally Shift Models Toward Lay Baselines</u>**

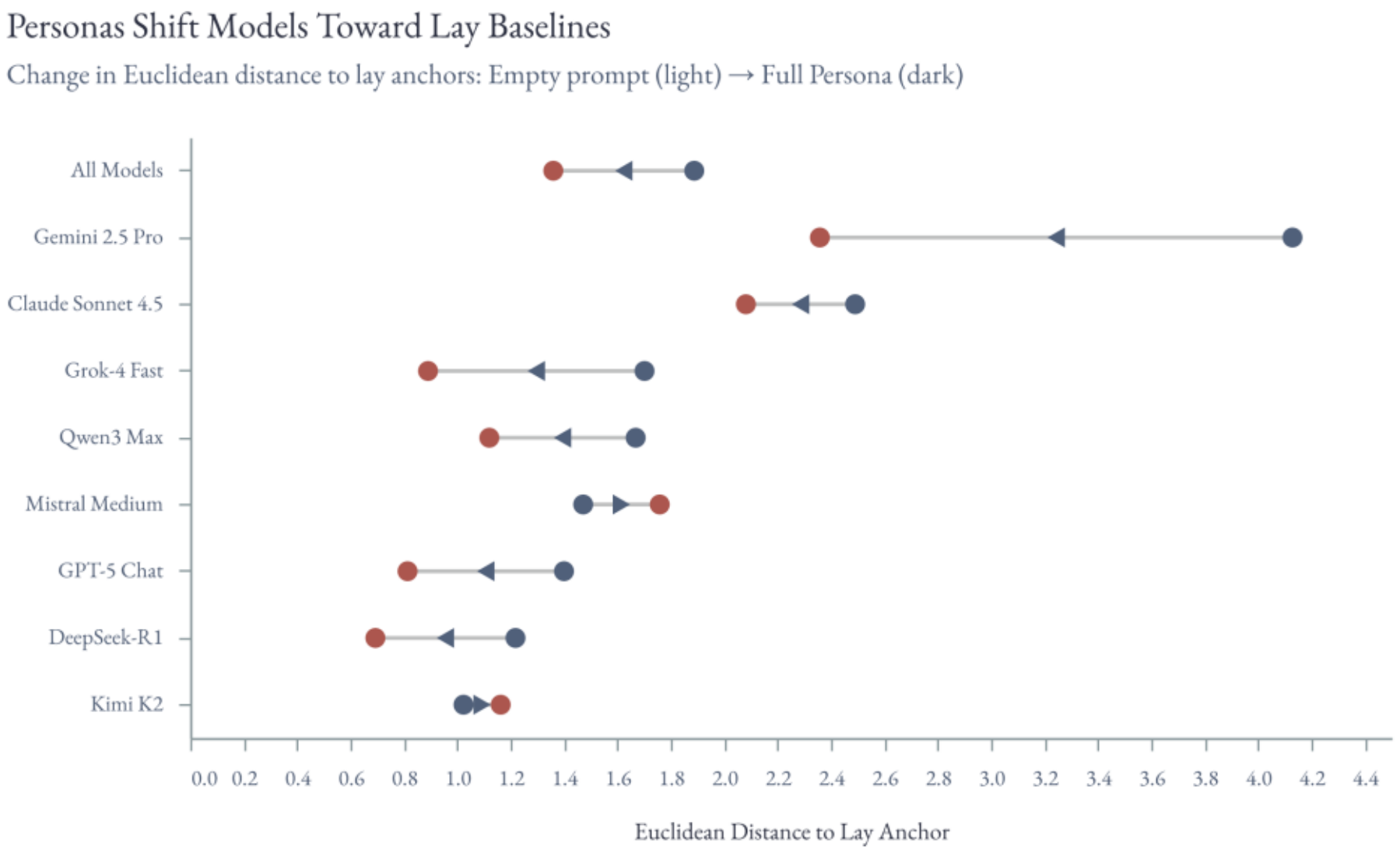

Averaging across models, persona prompts pull the mean model about 0.3 points closer to the lay baseline on each of consent, fairness, and legal obligation, and 6 of the 8 models move closer to lay subjects and farther from lawyers once personas are turned on.

The pattern is easiest to see if we put the three findings together. First, every model captures the formalist hierarchy that Sommers and Furth-Matzkin identify in lay contract reasoning: courts are expected to enforce more than people feel they consented to, and consent is more robust than fairness. Second, when we ask "who does this sound like" in a quantitative sense, most models land nearer to lay subjects than to elite legal professionals. Third, persona prompts push the models a little further into the lay corner of this space. For this kind of contract scenario, generative reasonable people do not, by default, talk like law school graduates but more like the consumers Sommers and Furth-Matzkin actually surveyed, and with the right persona scaffolding they sound slightly more like them still.

---

[136] A paired t-test across the eight architectures indicates that persona prompting significantly reduces Euclidean distance to the lay baseline, $t(7) \approx 2.4$, $p \approx 0.05$.

[137] Sign test, $p = 0.145$.

[138] (paired $t(7) = 2.38$, $p = .049$, $d = 0.84$)

# 4. THE PATH FORWARD

When Judge Kimba Wood was deciding Leonard v. Pepsico, she was engaging in the sort of humdrum reasoning judges are asked to perform almost every day: would any reasonable, perhaps even would *any*, person view a television ad promising a harrier jet for Pepsi points as a serious promise? [139] The Judge was not seeking to balance *b* against some *pl*, nor was she trying to school consumers on media literacy.[140] Rather than crafting a normative ideal, what she was actually trying to *discern* was whether a "genuine issue" existed:[141] whether ordinary people, especially the impressionable teenagers Pepsi targeted with its fizzy bravado, might have been misled into thinking they could trade bottle caps for a $23 million jet that "sure beats the bus".[142] As she noted, almost warily, the law calls on her to make such decisions alone, "just as [s]he decides any factual issue in respect to which reasonable people cannot differ."[143]

But what tools does the law give a judge, perched on the bench, to read into the minds of teenagers?

Now, for the first time, we have a tool to assist with such determinations. Before reaching he judgment, a judge in Judge Wood's shoes can consult models on how they, if they were roleplaying the target audience, understand the advertisement. Done correctly, the Judge could learn important information. This does not mean Judge Wood was wrong; she may have had sound normative reasons for holding as she did. But it suggests that her empirical confidence, that *no* reasonable person could see things otherwise, may have outrun her evidence.

The studies replicated here show that models can serve as important adjuncts to judicial decision-making—and beyond. In the this Part I discuss use cases, best practices, and some important limitations of this new tool. Ultimately, as with any new tool, much more exploration, validation, and tool-building is required; and this study seeks to showcase the utility and relevance of such a field of inquiry.

## 1. Interpretation of the Findings

### 1. LLMs as Social Calculators

The first study, replicating Jaeger's work on lay perceptions of negligence, tested a question that has long divided tort theorists: when people judge whether someone acted unreasonably, do they ask "was this efficient?" or "was this normal?"[144] Economic theories of negligence, crystallized in the Hand Formula, suggest people should engage in the former calculation. Social theories suggest the opposite: people judge reasonableness by conformity with common practice, treating what most people do as a proxy for what one ought to do.

[139] Leonard v. PepsiCo, Inc., 88 F. Supp. 2d 116, 127 (S.D.N.Y. 1999), aff'd, 210 F.3d 88 (2d Cir. 2000).

[140] United States v. Carroll Towing Co., 159 F.2d 169 (2d. Cir. 1947)

[141] Indeed, the logic in many contract interpretation cases is discerning the *actual* intent of the parties, rather than some ideal, value-maximizing judgment by an outsider. Alan Schwartz & Robert E. Scott, *Contract Theory and the Limits of Contract Law*, 113 YALE L.J. 541, 568 (2003) ("There is a consensus among courts and commentators that the appropriate goal of contract interpretation is to have the enforcing court find the 'correct answer.'").

[142] Leonard v. PepsiCo, Inc., 88 F. Supp. 2d 116, 121 (S.D.N.Y. 1999), aff'd, 210 F.3d 88 (2d Cir. 2000).

[143] *Id.*

[144] *See supra* Part 3.2.

Jaeger's human subjects sided decisively with the social theorists. When told that 90% of people took a particular precaution, they judged failure to take it far more harshly than when only 10% did so. The economic cost of the precaution barely registered.

The models replicated this hierarchy. Across architectures, the social-norm manipulation moved negligence judgments substantially more than the economic manipulation. Like their human counterparts, models appeared to have learned that what others do matters more than what efficiency demands.

What makes this finding notable is its resistance to doctrinal pedagogy. A model that had merely memorized legal rules, i.e., a stochastic parrot reciting case holdings, would predict the opposite pattern. Treatises and casebooks emphasize cost-benefit analysis; they treat custom as evidence of reasonableness but insist that "what ought to be done is fixed by a standard of reasonable prudence, whether it is usually complied with or not."[145] Yet models, like humans, inverted this doctrinal hierarchy. They learned from how people actually discuss and judge carelessness in the wild, not from prescriptive legal theory. In this sense, the models' training on natural discourse may give them *better* access to lay reasoning than a Yale-educated judge whose intuitions have been shaped by three years of doctrinal instruction and decades of professional socialization.[146]

### 2. *LLMs as Consent Evaluators*

The second study, replicating Sommers' work on consent under deception, presented a sharper test.[147] Sommers documented a paradox: people perceive more consent when someone is deceived by a material lie (about terms that matter to them) than by an essential lie (about the fundamental nature of what they're agreeing to)—even while acknowledging that material lies would matter more to the victim.13

This pattern defies both legal doctrine and intuition. Doctrine treats misrepresentations about value as more corrosive of consent precisely because they affect what the victim cares about. Yet human subjects showed the opposite, treating authenticity about a transaction's fundamental nature as more constitutive of consent than accuracy about its terms.

Models replicated this paradox with striking consistency. Seven of eight architectures reproduced both effects: material lies rated as mattering more, yet producing more perceived consent. Again, a model parroting doctrine would predict the opposite. That models captured this counterintuitive folk schema suggests they have internalized something deeper than legal rules—a lay theory of consent that legal professionals might not even recognize they diverge from.

### 3. *LLMs as Lay Judgment Estimators*

The first two studies asked whether models shift in the same direction as humans when experimental conditions vary. This sidesteps a thorny calibration problem: we need not ask whether a model's "6 out of 10" means the same as a human's, only whether both move upward under the same conditions. But practitioners will inevitably pose a harder question: when a generative reasonable person speaks, whose voice emerges? Does the model sound like a consumer scanning labels at the grocery store, or like a law professor parsing unconscionability doctrine?

---

[145] The T.J. Hooper, 60 F.2d 737, 740 (2d Cir. 1932)

[146] This observation parallels arguments in corpus linguistics that natural language usage, rather than expert intuition, should guide ordinary meaning analysis. *See Lee & Mouritsen, supra* note 7.

[147] *See supra* Part 3.3.

The third study addressed this directly.[148] Because Sommers and Furth-Matzkin fielded identical questions to both lay individuals and legal professionals from elite law schools, it provides two benchmarks for the same judgments. This enables a different test: not just whether models reason like humans, but which humans they resemble.

The findings suggest models, at least when personified, speak in lay registers. Every architecture replicated the structural hierarchy characterizing lay contract reasoning: hidden fees were deemed unfair, consent was judged as middling, and enforceability as high. This "formalist" pattern—where people believe courts will enforce terms they didn't truly consent to and find unfair—held across all eight models. Critically, when we measured distance to the two human baselines, most models landed closer to lay subjects than to legal professionals.

This finding addresses a natural concern about using AI as an empirical input: that it might launder professional intuitions under the guise of public sentiment. If models had absorbed primarily the skeptical, doctrine-informed perspective of legal elites, they would be of limited use in checking judicial assumptions against lay understanding. That they instead track ordinary consumers suggests they can serve the function this Article envisions.

### 4. *Synthesis: The Geometry of Reasonableness*

Taken together, the three studies suggest that certain LLMs have internalized an internal "geometry" of reasonableness judgments, not just isolated responses to individual prompts, but structural relationships between factors that mirror the architecture of human moral intuition. In the negligence setting, they track the intuitive priority of what people around us do over what a Hand Formula would prescribe. In the consent study, they reproduce a paradoxical structure in which lies that matter more to victims erode consent less. In the contract vignette, they mirror the hierarchy in which enforceability of deceptive contractual tactics feels sturdier than consent, and consent sturdier than fairness. In each domain, the models converge on a pattern that is recognizably human, even where it runs against legal orthodoxy.

Several features of this internalization deserve emphasis. First, it is consistent across domains. It seems that the same underlying capability that enabled models to prioritize social norms in negligence judgments also enabled them to replicate the consent paradox and to capture lay formalism in contract interpretation. This consistency suggests a generalizable capacity rather than domain-specific accident.

Second, it is robust across architectures. Eight models spanning proprietary and open-source systems, different training approaches, and varying parameter counts all captured the core directional patterns.

Third, and importantly for legal applications, models aligned more closely with lay intuitions than with doctrinal orthodoxy. While the fact that models are trained on large repositories of low quality online data (think comment sections or social media) is often seen as a limitation, garbage-in garabage-out, this may actually be a virtue. Models have seen how humans reason in the wild. A judge seeking to understand how "ordinary consumers" would interpret a disclosure, or how a "reasonable person" would perceive a police encounter, may find that a model trained on billions of words of natural human discourse has better access to that perspective than the judge's own professionally deformed intuitions.

---

[148] *See supra* Part 3.3.

Lastly, a critical finding is that while models showed an unexpected ability to detect the statistical patterns that drive reasonableness judgments, they struggled with calibration, the precise numbers they assigned. Models got the *direction* right: social norms matter more than economics, material lies produce more perceived consent than essential ones, lay people are contract formalists. But, as previous research shows, models assign different numerical scores and, as this study shows, they often amplified or compressed effects relative to human baselines.[149]

In the negligence study, models showed even stronger sensitivity to social norms than human subjects did, and detected economic effects that humans showed only as a statistical trend. In the consent study, models compressed their consent ratings toward the middle of the scale while clustering importance ratings at the ceiling. These patterns likely reflect the layering of alignment training through reinforcement learning,[150] the processes designed to make models safer, more polite, and less discriminatory, that are built on top of statistical impressions of how people actually talk and argue. The result is that models are not raw public-opinion pollsters. They carry institutional overlays that distort raw magnitudes even as they preserve directional patterns.

This limitation can be framed simply: generative reasonable people are better at answering "what tends to matter" than "how much should we adjust." The distinction matters practically. A regulator choosing between two disclosure formats could determine which makes a product warning feel more salient to consumers. A plaintiff's attorney deciding whether to emphasize the defendant's deviation from industry custom or the low cost of the foregone precaution could determine which framing moves simulated jurors more. A judge who wants to know whether the advertisers use of a specific whether including a specific disclaimer made the advertisement more or less serious in the eyes of teenagers. By contrast, a policymaker who wants to know whether a disclosure raises average consent ratings from 3.2 to 3.8 on a seven-point scale may demand a level of precision that current models cannot reliably deliver. Generative reasonable people are well suited to map the qualitative structure of folk judgment. They are not yet precision instruments for quantitative forecasting.

Beyond this specific limitation, we also cannot completely rule out all demand effects or other leakage, and having tested many models, random chance is still a possibility. It is also the case that this study only studies certain features of reasonableness and we cannot rule out that these were learned while others were not. An even more wicked problem is that we take for granted that the underlying studies reflect human perception: but if there are unknown faults in them or if they fail to generalize, then such inference would be unwarranted.

---

[149] On the recurring gap between directional replication and distributional or magnitude fidelity in "silicon sampling," including scale compression, ceiling effects, and subgroup misestimation, *See Bisbee et al., supra* note 21 (published online) (warning that synthetic survey data can diverge from human distributions even when correlations look strong); *See Dominguez-Olmedo et al., supra* note 100 (documenting substantial prompt sensitivity and measurement instability in LLM survey responses); *See Sarstedt et al., supra* note 16 (collecting reliability threats and recommending validation and design safeguards).

[150] Future work would want to compare effects between base models and models that underwent post-training.

### 2. Domains of Application: Between Promise and Prudence

The findings from the three studies demonstrate that generative reasonable people can provide a meaningful proxy for lay judgments across different domains of reasonableness. This capability addresses the democratic tension identified earlier: while the law aspires to reflect lay understanding, it often lacks accessible tools to discern it.

To explore domains of applications, it is helpful to distinguish three zones in which reasonableness standards operate:

***Explicitly Descriptive Standards.*** Some legal tests genuinely ask what ordinary people think, believe, or understand. Consumer protection standards assessing whether an advertisement would "mislead a reasonable consumer" fall here,[151] as do "ordinary meaning" inquiries in statutory interpretation and the "unsophisticated consumer" test in debt collection.[152] In these domains, the legal question is the empirical question. Silicon reasonable people offer the most direct value here, and their use raises the fewest legitimacy concerns.

***Explicitly Normative Standards.*** Some reasonableness tests are normative by design. The Hand Formula asks what precautions are cost-justified, a question of efficiency rather than social expectation. [153] Constitutional reasonableness tests often embed substantive values that override majoritarian preferences.[154] In these domains, silicon reasonable people are least appropriate as arbiters—but they retain value as transparency devices. If a court rules that "no reasonable person" would perceive a search as coercive, while simulated personas calibrated to the relevant community would perceive it as coercive, the court could profitably acknowledge that its ruling reflects a normative choice rather than an empirical finding

---

[151] See Federal Trade Comm'n, *Policy Statement on Deception* (Oct. 14, 1983), reprinted in *In re Cliffdale Assocs., Inc.*, 103 F.T.C. 110, 174–84 (1984) (defining deception by likely effect on consumers acting reasonably under the circumstances, and emphasizing audience-targeting); *Williams v. Gerber Prods. Co.*, 552 F.3d 934, 938 (9th Cir. 2008) (applying "reasonable consumer" standard in consumer deception context); *Fink v. Time Warner Cable, Inc.*, 714 F.3d 739, 741 (2d Cir. 2013) (reasonable consumer standard; plausibility-based dismissal where deception theory is implausible).

[152] For ordinary-meaning interpretation framed as the understanding of ordinary speakers, *See Holmes, supra* note 33, at 417; *See Tobia et al., supra* note 6 (survey-based approach to "ordinary meaning"). For the "least sophisticated consumer" (or closely related "unsophisticated consumer") framework in FDCPA litigation, see *Clomon v. Jackson*, 988 F.2d 1314, 1318–20 (2d Cir. 1993); *Avila v. Riexinger & Assocs., LLC*, 817 F.3d 72, 75–76 (2d Cir. 2016); see also *Gammon v. GC Servs. Ltd. P'ship*, 27 F.3d 1254, 1257 (7th Cir. 1994) (similar consumer-protective lens).

[153] See *United States v. Carroll Towing Co.*, 159 F.2d 169, 173 (2d Cir. 1947) (Hand, J.) (canonical formulation of negligence as a function of burden, probability, and loss, often rendered B < PL); Richard A. Posner, *A Theory of Negligence*, 1 **J. Legal Stud.** 29 (1972) (economic account of negligence as cost-justified precaution); Restatement (Third) of Torts: Liability for Physical & Emotional Harm § 3 (Am. L. Inst. 2010) (reasonable care standard keyed to foreseeable risks and precaution burdens); *See Gilles, supra* note 115 (mapping Hand formula's role and limits in doctrine and jury practice).

[154] See *Camara v. Mun. Court*, 387 U.S. 523, 536–37 (1967) (Fourth Amendment reasonableness as balancing, not a referendum on popular views); *Tennessee v. Garner*, 471 U.S. 1, 8–9 (1985) (deadly-force reasonableness framed as interest-balancing)

***Hybrid Standards***. Many reasonableness inquiries blend empirical and normative components.[155] Negligence asks what a reasonable person would do, but also what they should do; the two need not coincide. Consent inquiries ask both what people understood and what they were entitled to rely upon.[156] Contract interpretation asks what parties meant, but also what they should be held to.[157] In these domains, silicon reasonable people provide essential empirical input while leaving normative judgment to human decision-makers. They answer the descriptive predicate without resolving the prescriptive conclusion.

With those preliminaries in mind, let us now explore several domains of application and promise.

### 1. Lawmakers and Regulators: Scaling Public Voice

Policymakers wrestle with crafting rules that align policy goals with public perceptions. Consider the Federal Trade Commission's task of curbing unfair or deceptive practices.[158] The agency sought, in 2020, to conduct a periodical review of its rules on "made in the USA" claims made by sellers. A central challenge was to understand how consumers today interpret such claims: do they expect that virtually every part of the product was produced and assembled on the mainland? Or might they expect that key components or the majority of the value be produced in the USA?

To guide their rulemaking, the FTC relied on a survey of public perceptions: problem was, it was a quarter century old. In a globalizing world, attitudes *may* have shifted considerably since then.[159] The agency, which undertook a full revision of its rules, would certainly have benefitted from having a more contemporary understanding, but budget constraints limited their ability to acquire this information.[160]

This problem applies more generally to our methods of learning about public perceptions. Consumer surveys—perhaps the ideal form of feedback—are typically commissioned only for high-profile rulemakings due to cost constraints.[161] Other methods are also quite limited: Public comment periods overamplify mobilized voices, are subject to astroturfing, and are sometimes fall prey to mass form submission attacks.[162] Courts further complicate matters by oscillating between two interpretive poles: in some debt collection cases, they use the "least sophisticated consumer" standard,[163] and in others the "reasonable consumer."[164]

---

[155] For an argument in favor of the hybrid approach, *See Tobia, supra* note 9, at 343–45

[156] See Restatement (Second) of Torts § 892A cmt. c (Am. L. Inst. 1979) (consent as willingness in fact); id. § 892B(2) (consent induced by misrepresentation as to the essential character of the conduct is ineffective); Restatement (Second) of Contracts § 164 (Am. L. Inst. 1981) (fraudulent or material misrepresentation renders a contract voidable, subject to reliance and justification limits); *See Keeton et al., supra* note 9, § 18 (collecting doctrine on consent and invalidation through fraud/misrepresentation).

[157] *Lucy v. Zehmer*, 196 Va. 493, 503, 84 S.E.2d 516, 522 (1954) (objective manifestations control over undisclosed intent); *See Schwartz & Scott, supra* note 141, at 568 (framing interpretation as an attempt to reach the "correct answer," tempered by institutional limits).

[158] Federal Trade Commission Act, 15 U.S.C. §§ 41–58 (2018), see also N.Y. Gen. Bus. Law § 349 (McKinney 2020); Cal. Bus. & Prof. Code § 17200 (West 2020), Fla. Stat. Ann. § 501.201 et seq. (West 2020); Tex. Bus. & Com. Code Ann. § 17.46 (West 2020).

[159] Federal Trade Commission, *Made in the USA: An FTC Workshop*, BUREAU OF CONSUMER PROTECTION STAFF REPORT (June 2020). In addition, the FTC relied on a survey provided by Mark Hanna, the chief marketing officer of a US Jewler.

[160] *Id.*

[161] On these issues, *see supra* note 16 and accompanying text.

[162] Danielle A. Schulkin, *Improving the Management of Public Comments in a Digital Age*, **REG. REV. (NOV. 8, 2021)** ("comment process is susceptible to "astroturfing." . . . In some recent high-profile rulemakings,

Generative reasonable people offer a way out of this impasse. They present a handy mode of estimating public understandings, with little expense. A regulator could then sample test how models understand a claim like "made in the USA," and validate—at least [165] internally—whether there is a need to commission a full study. To a more limited degree of reliability, the models can engage in a "personaified" mode, roleyplaying personas—a harried parent scanning grocery shelves, a health-conscious gym-goer, a senior citizen with limited technical literacy—to provide perspective that drafters might miss and democratize insights at scale. Thus, to understand how a proposed notice might affect a reasonable consumer relative to an unsophisticated one, the agency could conjure different personas. Unlike sluggish notice-and-comment procedures, which may take months while generating skewed feedback, and unlike expensive surveys that require broad expertise in survey methodology, this method provides rapid, diverse perspectives at minimal cost.[166]

To illustrate, imagine a regulatory agency aiming to assess whether the "made from natural fruit" label on a juice product misleads consumers, given that the juice contains artificial flavors. The agency could deploy silicon personas in a robust RCT with two distinct conditions: in Condition A, personas are shown only the front label stating "made from natural fruit"; in Condition B, personas would view the full packaging, including an ingredient list that reveals the presence of artificial flavors. After exposure, each persona responds to targeted questions: "Do you believe this juice contains only natural ingredients?" (Yes/No) and "How natural do you think this juice is?" (rated on a 1-5 scale). By comparing the responses between the two conditions, regulators can determine if the label alone creates false impressions about the juice's composition. For example, if personas in Condition A are significantly more likely to assume the juice is entirely natural compared to those in Condition B, who see the artificial flavors listed, this would demonstrate that the label is confusing. These findings provide regulators with precise, actionable data to adjust labeling standards, ensuring they reflect consumer understanding and reduce deception, all before committing to extensive real-world consumer research.

Of course, there are tradeoffs. The method is less precise than a full survey, and the precision and reliability falls the more granular the simulated demographic is. The most useful question is the *realistic* alternative: if the standard mode of operation is to avoid public consultation, or have one that is susceptible to being hijacked by commercial or political interests, than a agencies have received—or have appeared to receive—millions of comments, many of which were fake or manipulated. . . . [and] garner large numbers of similar or identical comments, frequently in response to calls to action by public interest and advocacy groups"); Michael Herz, *Fraudulent Malattributed Comments in Agency Rulemaking*, 42 Cardozo L. Rev. 1, at 2 (2020) ("millions of other filings in the net neutrality docket appear to be the product of fraud").

[163] Avila v. Riexinger & Associates, LLC, 817 F.3d 72 (2d Cir. 2016)

[164] Jason E. Tavernaro v. Pioneer Credit Recovery, Inc., 20 F.4th 1234 (10th Cir. 2022). The Supreme Court has acknowledged that this is an unresolved issue in Sheriff v. Gillie, 578 U.S. 31, 40 n.6 (2016) (J. Ginsburg).

[165] As a quick test, GPT 4.5 responded: "A seller assembles bicycles in California using frames imported from China and domestically sourced wheels and gears. Suppose you are about to buy bikes and see a label "made in the USA." Would you consider the seller's representation to be accurate or misleading? ... [response:] Misleading, high confidence." Chat conversation, screenshot on file with author.

[166] Models differ in costs, but even a leading model—such as GPT O1—will only charge $15 per million words of input and $60 per million words of output, and prices decline rapidly. https://openai.com/api/pricing/

generative reasonable person offer a robust compromise, that is not less precise. For higher stake cases, or for instances where minority groups are likely to be deeply affected, the case for actual consultation increases. Fortunately, integrating generative reasonable people into the process could free up resources to that end.

### 2. Grounded Judicial Intuitions: Empirical Guardrails for Discretion

Consider a judge deciding whether a genuine issue exists when a consumer claims to have been misled by a claim that a product contains "33% less sugar,"[167] or that batteries are "up to 50% longer lasting."[168] It is hard for judges to truly put themselves in consumer shoes, and when they attempt to do so, in earnest and with diligence, they still come under fire for the insularity of elite intuition.[169] Judges, shaped by education and professional isolation, inadvertently risk conflating their own perspectives with those of the broader polity.[170] This could explain why in Beccera v. Dr Pepper, the court reasoned that consumers who see the term "diet" on the label will not understand it to imply that it will help with weight management.[171]

This judicial promise hinges on methodological rigor. There are many models that a judge could select from, and an infinite number of ways to prompt them. This is why transparency becomes crucial, if models inform judicial reasoning (while never replacing it), litigants should have procedural rights to know which model was used and how it was prompted. Federal Rule of Evidence 706, which governs court-appointed expert witnesses, provides a potential framework for integrating this technology while preserving adversarial testing

Under such a framework, both parties could retain rights to challenge model selection, question prompt construction, and propose alternative formulations - preserving adversarial testing while incorporating this new empirical tool. This approach acknowledges that generative reasonable people are neither neutral nor infallible, but rather one perspective-generating mechanism among many.

By treating generative reasonable people as one empirical data point, closer to a sophisticated survey than definitive proof, courts can harness their insights while preserving the deliberative integrity of judicial reasoning. This approach resonates with what Kahan and colleagues term "cognitive illiberalism"

---

[167] *Danone, US, LLC v. Chobani, LLC*, 362 F. Supp. 3d 109, 120–23 (S.D.N.Y. 2019) (denying preliminary injunction in part, but finding plaintiff's survey a "persuasive extrinsic evidence" that "overwhelming percentage" of consumers misunderstood "33% less sugar" claim)

[168] Millam v. Energizer Brands, LLC, No. 23-55192, 2024 WL 3294883, at *1–3 (9th Cir. June 14, 2024) (mem.) (affirming dismissal; "*up to 50% longer*" was not a promise of typical performance and would not deceive reasonable consumers)

[169] Even judges are aware of this issue. *See* Koehn v. Delta Outsource Grp., Inc., 939 F.3d 863, 864 (7th Cir. 2019) ("[T]he federal judges who must decide [FDCPA] motions are not necessarily good proxies for the "unsophisticated consumers" protected by the FDCPA.'"

[170] Jessica Guarino, Nabilah Nathani & A. Bryan Endres, What the Judge Ate for Breakfast: Reasonable Consumer Challenges in Misleading Food Labeling Claims, 35 LOY. CONSUMER L. REV. 82, 132 (2023) ("When a judge decides to impose their own beliefs and rationale into making determinations of whether a reasonable consumer would find a label misleading, food labeling litigation outcomes become inconsistent and inaccurate. Judges, unlike majority of the population, are highly educated. This can result in discrepancies in the approaches in which labels are scrutinized.")

[171] Becerra v. Dr Pepper/Seven Up, Inc., 945 F.3d 1225, 1231 (9th Cir. 2019) ("the allegations in the complaint fail to sufficiently allege that reasonable consumers read the word "diet" in a soft drink's brand name to promise weight loss")

in Fourth Amendment jurisprudence, where judges' cultural cognition often diverges from broader social perceptions of reasonableness.[172]

### 3. Litigants and Access to Justice

For litigants, particularly those from marginalized communities, the implications of handy generative reasonable people extend beyond doctrinal refinement to questions of access and equality. Litigation often exacerbates resource disparities, with mock trials and consumer surveys remaining prohibitively expensive for many.[173] Generative reasonable people could provide affordable approximations of jury perceptions, particularly for under resourced litigants. A tenant being told by a landlord that they are responsible for expensive repairs because they were not caused by normal wear and tear, could test such assumptions against a panel of simulated reasonable people.

Here, too, caution is advised: especially in the hands of inexperienced litigants, generative reasonable people may seem to hold greater truths than they actually do. The limitations of this tool might be forgotten in the name of convenience or unhelpfully suppressed by commercial providers marketing "AI jury prediction" services. As with most useful tools, there is potential for harm if misused, and it may be necessary to develop ethical guidelines for their deployment in litigation—perhaps through court rules or professional responsibility standards.

### 4. Guiding Firm Behavior: Better Compliance

Compliance departments can integrate silicon people into their internal processes, to effectively detect errors in corporate processes. In implementing the "reasonable security procedures and practices" required by the California Consumer Privacy Act,[174] companies could use generative reasonable people to evaluate whether their data protection measures would be considered adequate by the average consumer. Alternatively, if marketing wants to start an advertising campaign for fur coats, silicon people can be used to verify that a reasonable person would understand that the offer is subject to house rules. For firms that are trying to avoid unnecessary litigation or the ire of nudniks, such testing could prove a useful step in their compliance process.

In drafting mass contracts, this methodology could assist in achieving a greater degree of precision. There are many reasons why contract offers and terms are left uncertain,[175] O'Gorman counts twelve, some of which include routine negligence and motivated-reasoning reasons to see how the opposing party might understand a term.[176] A responsible attorney aiming to prevent future legal accidents may be able to study the contract's reasonable implications during negotiation, perhaps feeding some plausible scenarios and seeing how they might be interpreted under the contract.[177]

---

[172] https://scholarship.law.upenn.edu/cgi/viewcontent.cgi?article=3546&context=faculty_scholarship ("hat likely did not occur to the Justices in themajority was the degree to which their own perceptions (not to mention the perceptions of those who would agree with them upon watching the tape) would be just as bound up with cultural, ideological, andother commitments that disposed them to see the facts in a particular way") at 897

[173] See *supra* notes **Error! Bookmark not defined.** and 8.

[174] CAL. CIV. CODE §§ 1798.100 (e).

[175] RESTATEMENT (SECOND) OF CONTRACTS § 33(1) (1981). *See also* § 362 cmt. a. ("If this minimum standard of certainty is not met, there is no contract at all.").

[176] *See* Daniel P. O'Gorman, *The Restatement (Second) of Contracts Reasonably Certain Terms Requirement: A Model of Neoclassical Contract Law and a Model of Confusion and Inconsistency*, 36 U. HAW. L. REV. 169, 200-202 (2014).

Rather than replacing focus groups or market testing, generative reasonable people could serve as a preliminary screening tool, identifying potentially problematic language or claims before investing in more expensive consumer research. This tiered approach would allow companies to refine their compliance strategies iteratively, potentially catching issues that might otherwise emerge only after costly litigation has begun.

### 5. Legal Debates between the Descriptive to the Normative Person

Reasonableness has never lacked for theorists. It has lacked for facts. For two centuries, scholars have argued over the reasonable person's nature: descriptive or normative, majoritarian or aspirational, empirical anchor or moral ideal. The familiar move in these debates is to treat "what ordinary people think" as either dispositive (for descriptivists), irrelevant (for normativists), or a kind of loose constraint (for hybrid accounts). But across the spectrum, the arguments tend to proceed as if the empirical baseline is either already known, or impossible to know in any disciplined way. That assumption has quietly organized the field.

It is worth emphasizing how much contemporary scholarship still leans on descriptive imagery, even after the realist critique. Doctrinally, judges still speak in that register when they say that "no reasonable person" could think otherwise, or when they invoke "ordinary meaning" and "reasonable consumers" as if those are empirical categories rather than rhetorical devices.[178] And when legal academics are asked, in the abstract, what should inform assessments of reasonableness, they largely endorse a hybrid: custom and ordinary practice, plus evaluative judgment about what is good or fair, with markedly less enthusiasm for efficiency as the governing frame.[179] Even in an era that is self-conscious about the socially constructed nature of legal standards, the idea that reasonableness must remain tethered, at least in part, to ordinary social understanding continues to exert gravitational pull.

Yet, the field's ability to deliver on the reasonable person's promise has been light. Jury trials, in theory, instantiate community norms, but in practice they are episodic, selective, strategically framed, and usually absent. Surveys can help, but they are slow and expensive, and their design choices invite predictable disputes about framing and manipulation. As a result, the "empirical" side of reasonableness has often been supplied by intuition, and intuition is exactly what the realist critique targeted: the risk that judges and scholars launder their own priors through the idiom of the reasonable person.

Generative reasonable people challenge that scarcity premise. If silicon sampling can, when properly structured, recover the direction and internal architecture of lay judgments, the reasonable person debates no longer have to be fought in a vacuum. The contest between descriptive, normative, and hybrid accounts has always been partly conceptual, but it has also been pragmatic: what would it even mean to "track community standards" at scale, with reproducibility, and without a litigation budget? Once that question has a plausible answer, some familiar theoretical positions look less like principled jurisprudence and more like second-best adaptations to institutional incapacity.

---

[177] Reasonable implications are important throughout the law of contracts, see, e.g., RESTATEMENT (SECOND) OF CONTRACTS § 211(3) (1981).

[178] See, e.g., *Leonard v. PepsiCo, Inc.*, 88 F. Supp. 2d 116, 131 (S.D.N.Y. 1999) ("no reasonable, objective person" would treat the ad as an offer), aff'd, 210 F.3d 88 (2d Cir. 2000); *Anderson v. Liberty Lobby, Inc.*, 477 U.S. 242, 248 (1986) (summary judgment proper where "no reasonable jury" could find for the nonmovant). On courts' tendency to treat "ordinary meaning" as an empirically accessible baseline, often mediated through dictionaries or related tools, *See Lee & Mouritsen, supra* note 7.

[179] Eric Martínez & Kevin Tobia, *What Do Law Professors Believe About Law and the Legal Academy?*, 112 GEO. L.J. 111, 137–39 (2023) (describing survey recruitment and the law-professor sample), 159 (tbl. 4).

Because it is easy to see what generative reasonable people offer to the descriptivist, let us consider what it has in store for the normativist position. Normativists have long been right about one point that gets lost in caricature: the law often must depart from lay intuition. Efficiency, equality, constitutional constraints, and minority protection regularly require courts and regulators to resist majoritarian sentiment. But normativism faces its own recurring difficulty, one that typically remains implicit. Normative standards still have to operate in the world. They must be legible enough to guide behavior, stable enough to coordinate expectations, and administrable enough to be applied without constant second-guessing. This is where descriptive baselines matter even for those who reject their authority. If a proposed refinement of the negligence standard, the consent doctrine, or a disclosure regime relies on distinctions that most people do not grasp, the reform may be normatively attractive yet functionally self-defeating.

Generative reasonable people create a way to stress-test that problem early. Scholars can pilot normative proposals against simulated public comprehension, not to decide whether the proposal is justified, but to see whether it is communicable and where it predictably fails. That is a different kind of "constraint" than majoritarian deference, but it is a real one. Most importantly, the method sharpens the honesty of normative argument. Reasonableness rhetoric often invites a subtle slide. Courts announce a result in the language of description, "no reasonable person would…," while doing prescriptive work that could have been defended explicitly as policy. When a usable empirical baseline is available, that slide becomes harder to sustain. A judge or scholar who insists on a result that conflicts with lay understanding can still defend it, but the defense must be candid: the choice is normative, not a report about what ordinary people think. The rhetoric of the reasonable person stops functioning as a camouflage for discretion and becomes what it ought to be, a point of contestation that must be justified.

## 6. Principles and Best Practices

These are early days, so it would be imprudent to provide a definitive list of rules for application. Nevertheless, we can identify certain principles and best practices that should guide the use of generative reasonable people. A successful framework must balance three core tensions: between empirical fidelity and normative judgment, between majoritarian patterns and minority perspectives, and between technological capability and democratic accountability. Drawing on this Article's findings and their theoretical implications, the following considerations offer a preliminary roadmap.

### 1. Silicon Models as Adjuncts, Not Arbiters

First and foremost, generative reasonable people should augment—never supplant—human judgment. Their value lies not in resolving disputes but in surfacing latent assumptions about reasonableness that shape legal outcomes. Judges might use LLMs to test whether their intuitive application of "ordinary meaning" or "community standards" aligns with statistically common understandings, much as corpus linguistics aids textual analysis. Yet final determinations must remain tethered to law's normative commitments.

This principle yields two practical corollaries. First, transparency protocols should govern any legal use of these tools. Courts employing LLM-derived insights should disclose the model, prompt, and persona parameters, enabling adversarial testing through rebuttal via alternative models or prompts. Second, decision-makers should practice confidence calibration, treating model outputs as Bayesian priors rather than conclusive evidence. In Leonard v. Pepsico, for instance, a judge might note: "While GPT-4 suggests 68% of teenagers would perceive the advertisement as an offer, this finding aligns poorly with contract doctrine's objectivity standard, warranting significant discounting."

### 2. Addressing Bias as a First-Order Legal Concern

Generative reasonable people inherently feature a majoritarian bent—a characteristic that offers both advantages and risks. On the positive side, this tendency counters elite judicial intuition with aggregated lay perspectives. Yet this majoritarianism risks entrenching what critical scholars term the "reasonable man's" hegemony—the exclusion of marginalized voices from reasonableness's conceptual core.[180] Feminist critiques of "objective" standards in discrimination cases have long revealed how unexamined majorities distort fairness.[181]

Several mitigation strategies deserve consideration. Legal actors should engage in adversarial persona testing, probing minority viewpoints by prompting models to simulate intersectional identities and contrasting those outputs with majority responses. Regular empirical validation of model predictions against actual community feedback or surveys should be conducted to verify that persona-driven simulations accurately capture nuanced minority perspectives.

Institutions should adopt formalized audits analogous to Title VII's disparate impact framework. These audits should quantitatively measure differences in model-generated judgments across personas representing various protected classes. Models should undergo routine bias stress tests, deliberately introducing scenarios that historically trigger stereotypes or biases to evaluate whether the model reinforces or mitigates such biases.

Practitioners of this method should publicly disclose persona definitions and testing protocols to allow external scrutiny and accountability, facilitating ongoing refinement of methods.

Perhaps most critically, practitioners must maintain meaningful engagement with real-world minority communities. Model-generated outputs rapidly lose accuracy when intersectional complexity increases, and therefore these tools must complement—not replace—direct interaction with marginalized voices.

### 3. Acknowledging the Limits of Mimesis

While LLMs capture broad patterns in reasonableness judgments, their statistical abstractions cannot replicate the phenomenological richness of lived experience. Models may identify that 70% of simulated jurors consider a hidden contract term unfair, but they cannot articulate the visceral distrust of institutions that animates such judgments. This limitation necessitates contextual grounding practices.

---

[180] *See* Dimock, *supra* note 77 and accompanying text.
[181] Martha Chamallas, F*eminist Constructions of Objectivity: Multiple Perspectives in Sexual and Racial Harassment Litigation*, 1 TEX. J. WOMEN & L. 95 (1992).

Triangulation provides one essential safeguard: in high-stakes contexts, model outputs should be validated against traditional methods like surveys and focus groups. The FTC, for example, might compare LLM predictions about consumer confusion with A/B testing of actual advertisements. Narrative elicitation offers another approach, using prompting techniques to generate explanatory rationales, then assessing their coherence with qualitative accounts of reasonableness. Comparing model-generated narratives with jury deliberation transcripts, for instance, might reveal both alignments and divergences in reasoning patterns.

### 4. Ensuring Dynamic Representation

Legal standards of reasonableness evolve, but LLMs' training data freeze societal norms at a historical moment. This creates a "democratic lag" where models reflect past majorities, not present ones. The challenge mirrors originalism's dilemma: Should 2025 negligence judgments rely on a model trained pre-#MeToo or pre-pandemic?

Adaptive measures can partially address this concern. Temporal tagging—deploying metadata indicating a model's knowledge cutoff date—enables users to adjust for subsequent cultural shifts. Domain awareness represents another important safeguard: practitioners should avoid deploying these tools in contexts where social attitudes are shifting rapidly. For some applications, regular retraining or fine-tuning of models may be necessary to maintain alignment with contemporary social norms.

## 5. CONCLUSION

For two centuries, the reasonable person has been law's most convenient fiction: a figure invoked constantly, defined never, and measured almost not at all. When Judge Wood declared that "no reasonable, objective person" would see Pepsi's ad as a serious offer, she was not reporting a finding.[182] She was placing a bet that her intuition about teenage consumers tracked reality closely enough that no reasonable juror could disagree. The law gave her no tools to check.

This Article has shown that such tools now exist. Across three replications and nearly 10,000 simulated judgments, large language models recovered the internal architecture of lay reasonableness judgments. They did so not by memorizing doctrine, but by absorbing the patterns of human discourse from which reasonableness intuitions emerge. Models learned that social conformity trumps economic efficiency in negligence, that lies about essence corrode consent more than lies about value, and that ordinary consumers are contract formalists who expect courts to enforce terms they never meaningfully accepted. In each case, models captured schemas that run counter to what legal treatises teach, suggesting they learned from how people actually reason rather than from what lawyers say they should.

The implication is not that judges should outsource judgment to algorithms. It is that the empirical predicate of reasonableness standards (what ordinary people actually think) is no longer a black box accessible only through intuition, expensive surveys, or the episodic lottery of jury trials. It can be measured, at scale, for a fraction of traditional costs.

This changes the terms of an old debate. Scholars have long argued over whether the reasonable person is descriptive or normative, majoritarian or aspirational. But much of that debate was shaped by a constraint that seemed immutable: that lay judgments were too expensive and slow to surface reliably. Positions hardened around what was tractable rather than what was right. With that constraint loosened, we can ask the question more honestly. When courts depart from lay understanding, is that departure justified, or is it elite intuition dressed in populist clothing?

The reasonable person standard has always promised democratic legitimacy: law speaking in the voice of those it governs. But what ordinary people think, how they experience the world, and what they mean by their words, has always been illegible to the state. Generative reasonable people help make real ordinary people visible.

---

[182] Leonard v. PepsiCo, Inc., 88 F. Supp. 2d 116, 127 (S.D.N.Y. 1999), aff'd, 210 F.3d 88 (2d Cir. 2000).